\documentclass[useAMS,usenatbib]{mn2e}
\usepackage{graphicx} 

\title[Multi-epoch infrared photometry of the star forming region G173.58+2.45]
{Multi-epoch infrared photometric study of the star forming region G173.58+2.45}
\author[W. P. Varricatt,  C. J. Davis \& A. J. Adamson]{Watson P. Varricatt$^{1}$\thanks{E-mail:
w.varricatt@jach.hawaii.edu(WPV); c.davis@jach.hawaii.edu(CJD); a.adamson@jach.hawaii.edu(AJA)},  Christopher J. 
Davis$^{1}$\footnotemark[1] \& Andrew J. Adamson$^{1}$\footnotemark[1] \\
$^{1}$Joint Astronomy Centre, 660 N. Aohoku Place, Hilo, HI-96720, USA}
\begin{document}

\date{Accepted Jan. 2005. Received ..; in original form 2004 Oct. 12}

\pagerange{\pageref{firstpage}--\pageref{lastpage}} \pubyear{2004}

\maketitle

\label{firstpage}

\begin{abstract}
We present a multi-epoch infrared photometric study of the intermediate-mass 
star forming region G173.58+2.45. Photometric observations are obtained using 
the near-infrared $JHKL'M'$ filters and narrow-band filters centered at the 
wavelengths of H$_2$ (1-0) S(1)  (2.122 $\mu$m) and [FeII] (1.644 $\mu$m) 
lines.  The H$_2$ image shows molecular emission from shocked gas, implying 
the presence of multiple star formation and associated outflow activity.  We 
see evidence for several collimated outflows.  The most extended jet is at 
least 0.25 pc in length and has a collimation factor of $\sim$ 10, which
may be associated with a binary system within the central cluster,
resolved for the first time here.  This outflow is found to be 
episodic; probably occurring or getting enhanced during the periastron 
passage of the binary.  We also find that the variable star in the vicinity 
of the outflow source, which was known as a FU Ori type star,  is probably 
not a FU Ori object.  However, it does drive a spectacular outflow and the 
variability is likely to be related to accretion, when large clouds of 
gas and dust spiral in towards the central source. Many other convincing
accretion-outflow systems and YSO candidates are
discovered in the field.

\end{abstract}

\begin{keywords}
infrared: stars -- stars: formation -- stars: colours -- binaries: visual -- 
circumstellar matter -- ISM: Jets and Outflows -- ISM: individual: G173.58+2.45
\end{keywords}

\section{Introduction}
It is now known that massive and intermediate-mass Young Stellar Objects (YSOs)
drive bipolar outflows similar to their low-mass counterparts, albeit on larger 
scales (Shepherd and Churchwell 1996; Churchwell 1997; Cesaroni et al. 1997; 
Kumar, Bachiller \& Davis 2002;  Davis et al. 2004;  Beuther, Schilke and 
Gueth 2004). Massive star formation occurs in clusters and is often associated 
with lower mass star formation.  In this paper, we discuss the case of an 
intermediate-mass YSO G173.58+2.45, where multiple star formation is taking 
place within a cluster.  

G173.58+2.45 (IRAS 05361+3539) is embedded in the centre of a large molecular
cloud. Shepherd \& Churchwell (1996) noticed that the IRAS fluxes from the
embedded source meet the criteria for an Ultra Compact HII region (UCHII)
(Wood \& Churchwell 1989). From CO observations, they estimated a 
line-of-sight velocity with respect to local standard of rest, $v_{LSR}$, of
-18 km s$^{-1}$, which is very close to the value of -17 km s$^{-1}$ estimated
by Bronfman, Nyman \& May (1996).  Wouterloot et al. (1988) and Palagi et al.
 (1993) detected H$_{2}$O maser emission from this field. Zinchenko, Pirogov 
\& Toriseva (1998) detected a CS($J$=2-1) core, which peaks 
$\sim$ 10$\arcsec$ east of the IRAS position.  The kinematic distance to the 
source was estimated by  Wouterloot \& Brand (1989) to be 1.8 kpc.

This region was studied at millimeter wavelengths (in $^{12}$CO and $^{13}$CO)
by Shepherd \& Churchwell (1996). Their $^{12}$CO spectra 
showed high velocity line wings indicating the presence of CO outflows. They
identified it as a bipolar molecular outflow candidate and estimated the
mass in the outflow to be 32 M$_{\odot}$. Shepherd and Watson (2002) (SW02 
hereafter) conducted detailed observations of the region at radio, mm and 
infrared wavelengths.  CO emission line observations carried out by them 
detected a large scale outflow extending 3$\arcmin$.4 end-to-end,
which is 1.8 pc at a distance of 1.8 kpc.  At 3 mm, they also detected two 
continuum emission sources, which are associated with 
reddened infrared sources.  They proposed that the outflow in this field
was driven by at least two intermediate-mass YSOs.

Very few near-infrared (IR) studies of this region have been done.  The H$_2$ 
image of {\citet{coar00}} detected signs of nebulosity in the direction of 
the outflow proposed by Shepherd and Churchwell (1996).  Later, more 
sensitive IR observations by SW02 failed to detect any signs of the jet
proposed by {\citet{coar00}, instead, resolved it to be a chance alignment of 
IR sources.  However, the existing near-IR observations lack the sensitivity 
and spatial resolution required and hence, this study.

\section[]{Observations}

Observations were acquired with the 3.8m United Kingdom Infrared Telescope 
(UKIRT), Mauna Kea, Hawaii. Photometry in the $J$ and $K$ bands were obtained
on 2001 Dec., 26 using UFTI (Roche et al. 2002) equipped with a Rockwell
1024$\times$1024 HgCdTe array, giving a pixel scale of 0$\arcsec$.091/pixel
and a field of view of 1$\arcmin$.5$\times$1$\arcmin$.5 at the f/36 
cassegrain focus of UKIRT. 
The region was again observed using UFTI on
2002 Oct., 22 in the $H$ and $K$ bands and in narrow-band filters at the
wavelengths of the H$_{2}$ {\it{v}}=1-0 S(1) line (2.1218 $\mu$m) and the 
[FeII] $a^{4}D_{7/2} - a^{4}F_{9/2}$ line (1.6439 $\mu$m) and on 6 Nov 2002, 
in the $J$ band. The observations were obtained by jittering the telescope 
on 9 points with 20$\arcsec$ offsets from the centre resulting in a 
field of view of 2$\arcmin$.2$\times$2$\arcmin$.2 for the mosaics.
The narrow-band
H$_{2}$ and [FeII] filters have FWHM bandwidths of 0.02 $\mu$m  and
0.016 $\mu$m respectively. The median seeing was $\sim$0$\arcsec$.6 on 
2001 Dec. 26 and $<$ 0$\arcsec$.4 on 2002 Oct. 22 and on Nov. 6. The 
better seeing on the latter days enabled us to achieve excellent spatial
resolution and also to
detect many faint H$_{2}$ emission features. On 2003 March 19, the field was
again observed in $JHK$ using UKIRT and UIST (Ramsay Howat et al. 2000),
which uses a 1024$\times$1024 InSb array, at an image scale of 
0$\arcsec$.12/pixel. A 9 point jitter with offsets of 30$\arcsec$ 
from the centre gives a field of view of 3$\arcmin$$\times$3$\arcmin$ for the 
UIST images.  $JHK$ photometric standard stars from the list of UKIRT faint 
standards close to the object were always observed prior to the object in 
each filter. All observations were obtained under photometric sky conditions 
at airmass less than two.
The observed magnitudes were corrected for the small difference in airmass 
between the object and the standard using the average extinction/airmass for
Mauna Kea for each filter.  Observations were again performed in the $L'$ and 
$M'$ bands using UKIRT and UIST on 2003 March 13 and Dec. 23. The observations
were carried out by jittering the image on four positions on the array and 
taking the difference of adjacent jittered frames after flat fielding, since 
the background radiation is not steady at these wavelengths. On March 13, 
both the $L'$ and the $M'$ band observations were acquired using the 
512$\times$512 sub-array and on Dec. 23rd, the $L'$ band observations were 
carried out using the 1K$\times$1K array and $M'$, using the 512$\times$512 
sub-array.  $L'$ and $M'$ photometric standards 
close to the object, from the list of Leggett et al. (2003) for the MKO-NIR 
system, were observed for photometric calibration.  Table {\ref{obslog}} shows 
the log of observations.

\begin{table}
\caption{Log of photometric observations}
\label{obslog}
\begin{minipage}{140mm}
\begin{tabular}{lllll}
\hline
Filter    &Date          &Int. time &Instrument &Field of  \\
          &yyyymmdd      &(sec)     &           &view   \\
\hline
$J$       &20011226      &540       &UFTI       &2$\arcmin$.2$\times$2$\arcmin$.2 \\
          &20021106      &675       &UFTI       &2$\arcmin$.2$\times$2$\arcmin$.2 \\
          &20030319      &90        &UIST       &3$\arcmin$.0$\times$3$\arcmin$.0 \\
$H$       &20021022      &810       &UFTI       &2$\arcmin$.2$\times$2$\arcmin$.2 \\
          &20030319      &270       &UIST       &3$\arcmin$.0$\times$3$\arcmin$.0 \\
$K$       &20011226      &360       &UFTI       &2$\arcmin$.2$\times$2$\arcmin$.2 \\
          &20021022      &1080      &UFTI       &2$\arcmin$.2$\times$2$\arcmin$.2 \\
          &20030319      &270       &UIST       &3$\arcmin$.0$\times$3$\arcmin$.0 \\
$L'$      &20030319      &104       &UIST$^*$   &1$\arcmin$.3$\times$1$\arcmin$.0 \\
          &20031223      &529       &UIST       &2$\arcmin$.7$\times$2$\arcmin$.0 \\
$M'$      &20030319      &115       &UIST$^*$   &1$\arcmin$.5$\times$1$\arcmin$.0  \\
          &20031223      &553       &UIST$^*$   &1$\arcmin$.5$\times$1$\arcmin$.0 \\
H$_2$     &20021022      &900       &UFTI       &2$\arcmin$.2$\times$2$\arcmin$.2 \\
$[$FeII$]$      &20021022      &900       &UFTI       &2$\arcmin$.2$\times$2$\arcmin$.2 \\
\hline
\end{tabular}
\end{minipage}\\
$^*$ observed using 512$\times$512 sub-array. All the rest were observed
using the full 1024$\times$1024 array.
\end{table}

Preliminary reduction of the data was carried out using ORACDR - the pipeline
data reduction facility available at UKIRT.  Dark frames were obtained in 
each filter at the start of each set of observations.  Sky flats were
generated by median combining the dark- and bias-subtracted
observed frames.  The flat fielded individual frames were finally combined 
to produce mosaics using the STARLINK packages CCDPACK and KAPPA,
by matching the centroids of bright stars in the overlapping fields in 
each pair of dithered frames. Fig. {\ref{g17358JHK}} shows the $JHK$ colour
combined image of the field observed on 2002 Oct 22 and Nov. 6.  The stars
appearing red in colour are highly embedded objects and those in blue are
foreground objects. Fig. {\ref{g173L}} shows the central region of the
$L'$ image observed on 2003 Dec 23. The $M'$ image is not shown here.

\begin{figure*}
\vspace{200pt}
\includegraphics{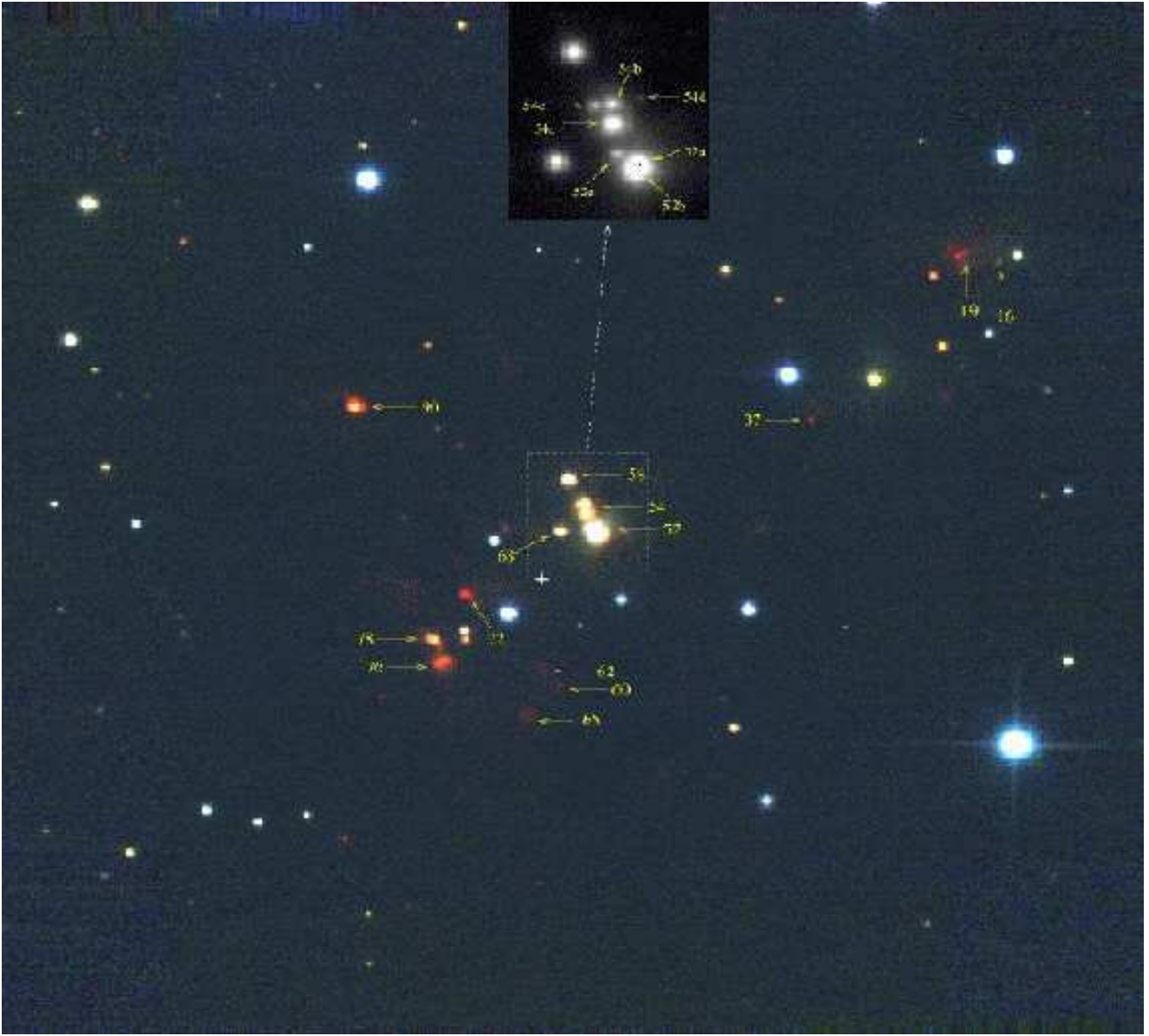}
\vspace*{310pt}
\caption{$JHK$ colour composite picture of the observed field constructed from
the $J$ band image observed on 2002 Nov. 11 and the $H$ and $K$ band images
observed on 2002 Oct. 22.  Most of the objects that appear blue in colour are
foreground stars and those that are yellow and red in colour are embedded in the
cloud. The red sources are extremely embedded objects;  many of them are YSOs
with IR excess.  The position of the IRAS source is marked with a '+'.
The central region is shown in the inset.  The peaks of \#52a and b are
marked in the figure with a single contour to show the binary components.
The orientation of the figure is North-up and East-left}
\label{g17358JHK}
\end{figure*}

\begin{table}
\caption{$L'$ and $M'$ photometric magnitudes.  Sources outside the field of
view are indicated; a dash indicates sources not detected.  For \#54, the $L'$
magnitudes of the components are given only for 20031223 when the S/N was
better. }
\label{LMPhot}
\begin{tabular}{@{}lrrrr}
\hline
No:   &\multicolumn{2}{c}{------- $L'$ Band -------}&\multicolumn{2}{c}{------- $M'$ Band -------} \\
        &20030319   &20031223   &20030319     &20031223    \\
\hline
                                                                                                                                                
14      &not obs  &9.92  .08  &\multicolumn{2}{c}{not obs} \\[.1mm]
15      &not obs  &12.29 .15  &\multicolumn{2}{c}{not obs} \\[.1mm]
19(a+b) &not obs  &11.55 .06  &\multicolumn{2}{c}{not obs} \\[.1mm]
25      &not obs  &12.35 .11  &\multicolumn{2}{c}{not obs} \\[.1mm]
31      &not obs  &11.98 .09  &\multicolumn{2}{c}{not obs} \\[.1mm]
32      &not obs  &12.43 .11  &\multicolumn{2}{c}{not obs} \\[.1mm]
37      &not obs  &12.92 .25  &\multicolumn{2}{c}{not obs} \\[.1mm]
38      &not obs  &12.31 .23  &\multicolumn{2}{c}{not obs} \\[.1mm]
41      &not obs  &13.36 .35  &\multicolumn{2}{c}{not obs} \\[.1mm]
52(a+b) &9.22  .03&9.34  .04  &8.39  .06 &8.49 .06         \\[.1mm]
54a     &         &10.48 .02  &          &                 \\[-.8mm]
54b     &         &11.98 .04  &          &                 \\[-.8mm]
54d     &         &12.56 .10  &          &                 \\[-.8mm]
54(a+b+d)&10.24 .04&10.11 .03 &9.50  .20 &9.47 .13         \\[.1mm]
58      &11.27 .03&11.46 .06  & --       & --              \\[.1mm]
59      &--       &12.54 .28  & --       & --              \\[.1mm]
60      &--       &detected   & --       & --              \\[.1mm]
62      &--       &detected   & --       & --              \\[.1mm]
67      &--       &12.73 .24  & --       & --              \\[.1mm]
68      &--       &12.54 .25  & --       & --              \\[.1mm]
72      &10.67 .05&10.78 .05  &9.89 .14  &9.76 .09         \\[.1mm]
78      &10.41 .10&10.66 .03  &8.51 .11  &8.79 .10         \\[.1mm]
85      &not obs  &11.24 .05  &\multicolumn{2}{c}{not obs} \\[.1mm]
90      &10.24 .03&11.09 .09  &9.29 .10  &10.14 .20        \\[.1mm]
119     &not obs  &11.59 .06  &\multicolumn{2}{c}{not obs} \\[.1mm]
120     &not obs  &12.10 .10  &\multicolumn{2}{c}{not obs} \\[.1mm]
\hline
\end{tabular}
\end{table}

The field is crowded in the $JHK$ bands. So, we estimated the magnitudes
in these bands using the DAOPHOT package of IRAF, by fitting a point spread
function (PSF) to the individual objects in the field. For each mosaic, the
PSF is generated by fitting profiles simultaneously to several isolated stars
in the field. Table {\ref{JHKPhot}} shows the $JHK$ magnitudes of the stars
detected on all the epochs of our observations.  For the $L'$ and
$M'$ images, since fewer objects are detected, aperture photometry was
done using the STARLINK display and data reduction package GAIA.
Table {\ref{LMPhot}} lists the $L'$ and $M'$ magnitudes estimated from
our photometry.

SW02 carried out astrometric calibration of the field. We have, therefore,
defined our coordinates to match theirs.  Coordinates of
the objects detected in our observations are calculated using the IRAF 
tasks CCMAP and CCTRAN, adopting the coordinates of some of the 
isolated stars from SW02 as reference. Our coordinates are West and North
of the 2MASS by 0.$\arcsec$9 and 0.$\arcsec$8 respectively. 
In Tables {\ref{JHKPhot} and {\ref{LMPhot},
we have also included the identifications of SW02 and Chakraborty et al (2000),
along with our new identification numbers,  for ease of comparison.
Many new objects are detected in our observations. Some
of these objects appear to be non stellar since they shine almost
equally bright in the H$_2$ images as in $K$. They are separately listed
in Table {\ref{hhflux}}. These are likely to be the molecular equivalents of
'Herbig-Haro' (HH) objects, produced by the outflows associated
with the multiple star formation activity in the region.

\begin{figure}
\vspace{154pt}
\includegraphics{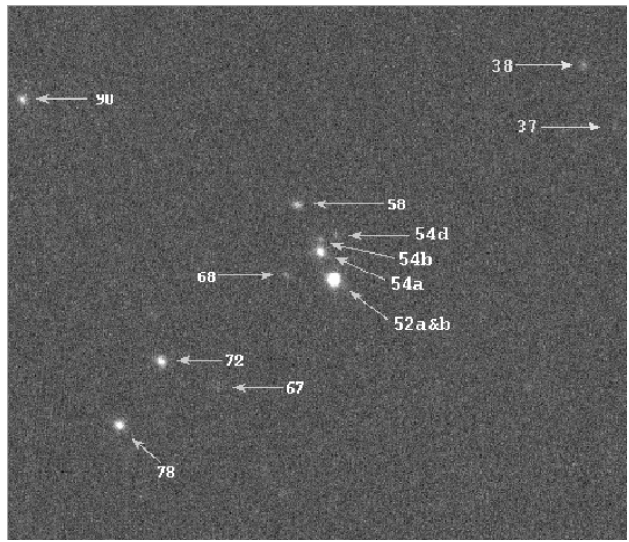}
\vspace{70pt}
\caption{$L'$ image of the central region observed on  23 Dec 2003 }
\label{g173L}
\end{figure}

The narrow-band H$_2$ and [FeII] images were continuum subtracted using the
scaled $K$ and $H$ band images respectively. For continuum subtraction, 
the integrated, sky subtracted counts on many isolated point sources (without 
any IR excess) were measured in the narrow-band filters and in the $H$ 
and $K$ filters using a circular aperture typically $\sim$4 times the 
FWHM.  The ratios of the counts $K$/H$_2$ and $H$/[FeII] were evaluated for 
these stars and the average values were taken.  The constant sky backgrounds
were subtracted from the 
$H$ and $K$ band images. These images were then divided by the above ratios 
and were subtracted from the sky subtracted narrow-band filter images.  
The diffuse continuum got subtracted out well. However, stars often gave 
residuals due to changes in PSF between the broad- and the narrow-band
observations.
Continuum subtracted H$_2$ images (Figs. \ref{g17358h2}, \ref{g17358region3}
and \ref{region2h2}) reveal several 'HH' type emission features and 
at least six bow shocks,  most of which are expected to 
be from shocked H$_2$. The continuum subtracted [FeII] image did not show 
any of these features except for one bow shock (B5). Therefore,
it is not shown here.

\section{Results and Discussion}

\subsection{$JHKL'M'$ Photometry}

The coordinates of the objects detected and their magnitudes estimated for 
all the epochs of our
observations are listed in Tables \ref{JHKPhot} and \ref{LMPhot}.
The objects detected in the $L'$ and $M'$ band imaging are the embedded stars 
in Fig. {\ref{g17358JHK}}, most of which show IR excess and a few of the 
brightest foreground stars.  The region is studied using colour-colour and 
colour-magnitude diagrams.

\subsubsection{Colour-colour and colour-magnitude diagrams}
                                                                                                                                                
The near IR colour-colour and colour-magnitude diagrams are useful tools
for identifying YSO candidates. Fig. {\ref{g173JHHK}} shows the stars plotted
on the $(J-H) - (H-K)$ colour-colour diagram.  Most of the data are from
the UFTI observations of 2002 Oct. 22 and Nov. 6, when the seeing was best.
Since the UIST mosaics cover a wider field, the stars outside the UFTI field
plotted in this figure are taken from the UIST observations of 2003 Mar. 19.
The continuous line shows the distribution of ($J-H$) against ($H-K$) of
un-reddened  main sequence stars and the dashed line, that of un-reddened giant
stars, constructed from the intrinsic colours of Koornneef (1983).  The lower
end of the curve marks the early type stars and the upper end, the late type
stars.  The diagonal parallel lines are the reddening vectors up to A$_V$=30,
at the maxima of the main-sequence and the giant stars' colours and at the
minimum of the main sequence colours. This is the direction in which reddening
moves the stars in the colour-colour diagrams.  We adopted a reddening law with
R$_V$=(A$_V$/E($B-V$))=5, which is typical for dense clouds (Cardelli, Clayton
\& Mathis 1989).  This gives values of extinction varying from A$_J$=9.82 to
A$_{L'}$=1.66 for $J$ to $L'$ at A$_V$=30. The stars in the reddening band,
the region between the
parallel lines, are main-sequence and giant stars at different reddening
within the cloud and along the line of sight.  The region below the lower line
is occupied by YSOs (Lada \& Adams 1992), where CTTS, WTTS,
HAeBe stars and luminous YSOs occupy different regimes. The
excellent spatial resolution has enabled us to resolve the multiplicity of
many objects, which were perceived as singles
in the previous investigations.  The colours of these objects measured
as singles and also of the resolved component stars measured individually
are plotted in the figure.

\begin{figure*}
\vspace{140pt}
\includegraphics{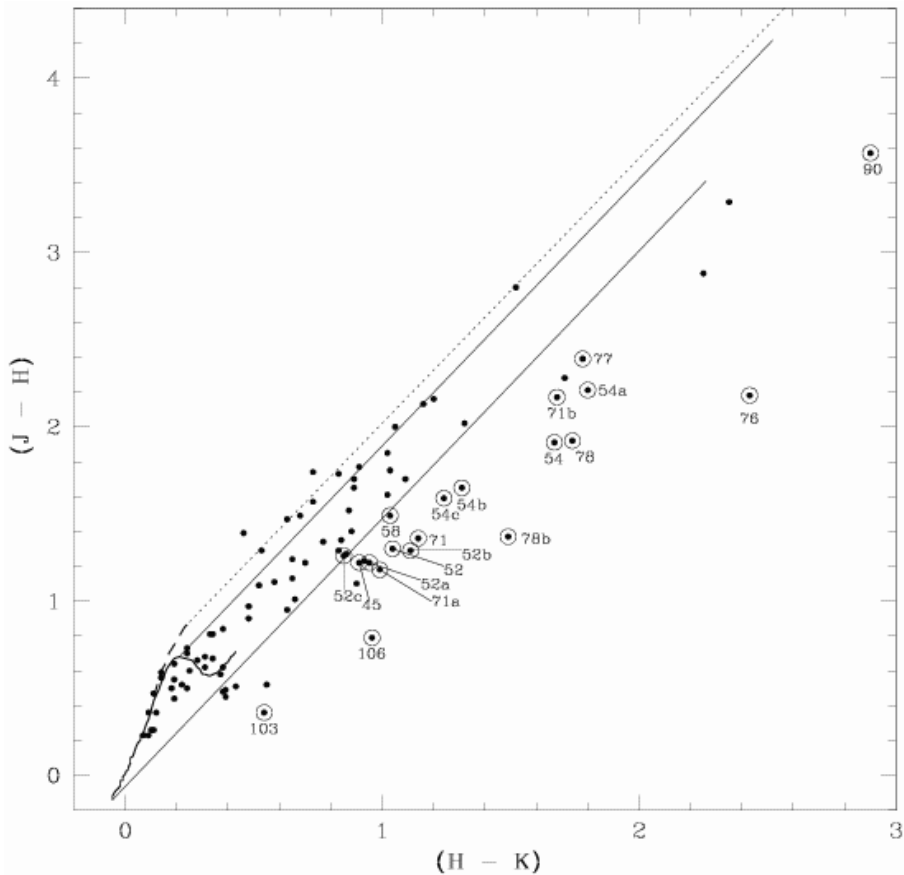}
\vspace*{200pt}
\caption{($J-H$) - ($H-K$) colour-colour diagram of the field derived from our
observations. The continuous line shows the locus of the un-reddened main sequence
stars and the dashed line, that of giant stars. The straight lines show the
reddening vectors up to A$_V$=30.}
\label{g173JHHK}
\end{figure*}

In Fig. {\ref{g173KHK}}, we have plotted the $K$ magnitudes of the objects
against their ($H-K$) colours. The continuous line is the location of the 
un-reddened main sequence stars from Koornneef (1983) at a distance of 1.8 kpc.
The location of the
different spectral types are marked on this line.  The dotted lines show the 
reddening vectors up to A$_V$=30.  As in Fig. {\ref{g173JHHK}}, the locations
of the individual components of multiples and that of the components put 
together are separately shown.  For reddened main sequence stars, 
extrapolation of the stars back along the reddening vector to the 
main-sequence line should give their spectral types.  However, many of the 
reddened stars, especially the YSOs,
will have infrared excess arising from circumstellar matter, which makes them
brighter towards longer wavelengths and thereby,  moves them up in the 
colour-magnitude diagram. So, great caution has to be taken when drawing 
conclusions about spectral types from the colour-magnitude diagram. For objects
with infrared excess, it should be treated only as an upper limit to the 
spectral type.

\begin{figure*}
\vspace{140pt}
\includegraphics{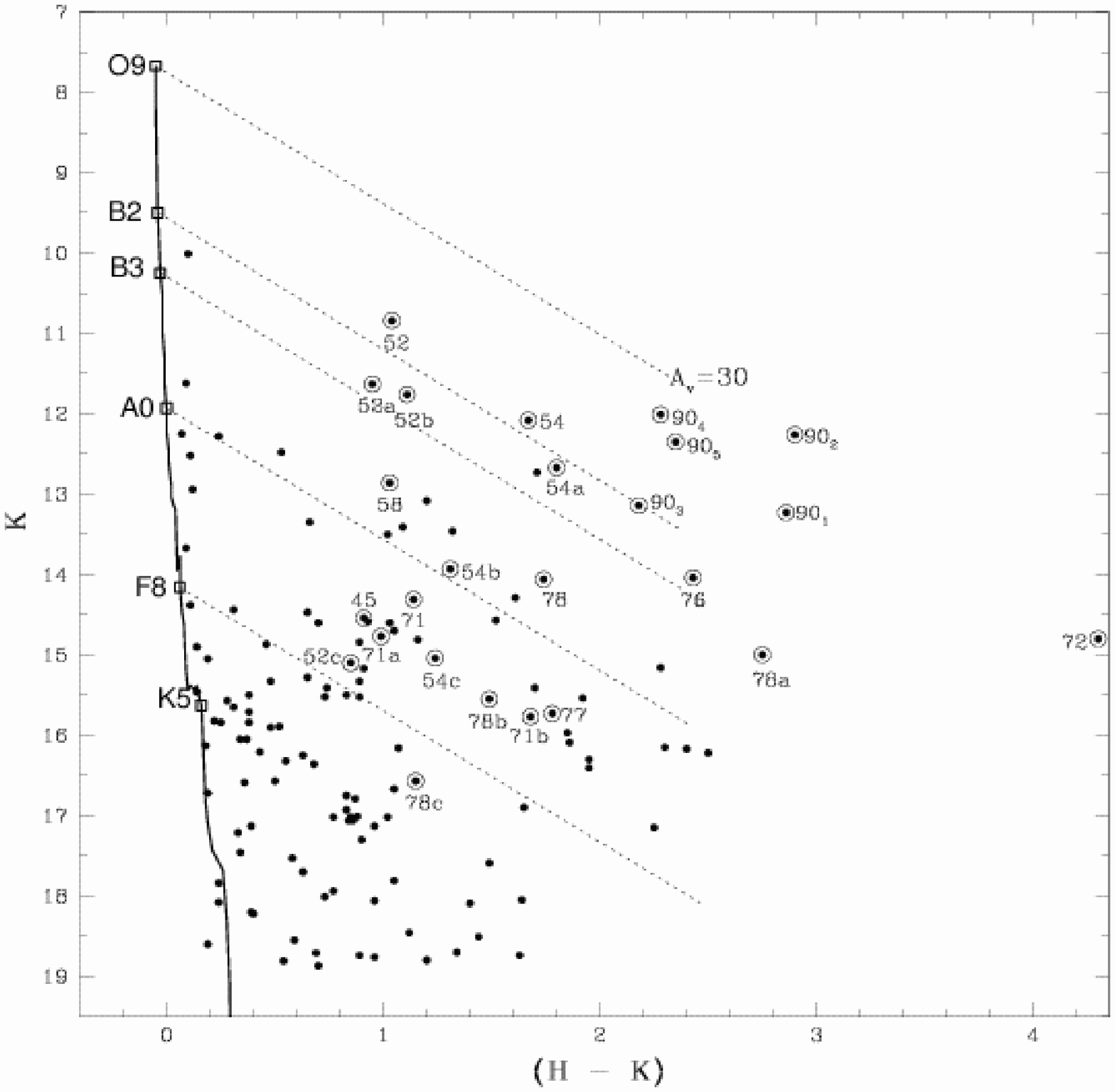}
\vspace*{200pt}
\caption{$K-(H-K)$ colour magnitude diagram.  The continuous line shows the
locus of main sequence stars at a distance of 1.8 kpc.  The straight lines show the
reddening vectors up to A$_V$=30. Stars showing IR excess are circled.}
\label{g173KHK}
\end{figure*}
Similarly, an ($H-K$) - ($K-L'$) colour-colour diagram (Fig. \ref{g173HKKL})
is constructed using the infrared colours for main sequence and giant stars
given by Bessell \& Brett (1998).  Most of the stars detected in the $L'$
band are those with IR excess in the $K$ band and a few bright
foreground stars.  The YSO candidates are enclosed in circles and are labelled
in the figure.  These objects, which show IR excess in Figs. \ref{g173JHHK} and
\ref{g173HKKL}, are discussed in section 3.3.

{\subsection {The narrow-band H$_2$ and [FeII] imaging}}
Narrow band imaging at the wavelength of the H$_2$ (1-0) S(1) line at 
2.122 $\mu$m is very useful for detecting shock excited emission from molecular
hydrogen and, thereby, the ongoing star formation and outflow activity in 
embedded 
regions.  Fig. {\ref{g17358h2}} shows the continuum-subtracted H$_2$ image 
observed using UFTI.  Many of the emission features detected are labelled in 
the figure and their coordinates and integrated fluxes are listed in Table 
{\ref{hhflux}}.  Arrows are drawn in the direction of three well aligned sets 
of emission knots; one set is probably produced by the bright source
(IRS1) located near the centre of the field.  Many of the H$_2$ emission 
features are seen to be associated with reddened objects in the field,
which are described below.  Some
features, which are enclosed in circles, are not positively traced back to any 
specific object in the field.  These features could be from some of the
objects described below or from embedded objects, which are not detected at
the current depth of integration.

We used an integration time of 900 seconds
in both the filters, which, in [FeII], gave a 3$\sigma$ sensitivity of
$\sim$ 2 $\times$ 10$^{-20}$ W m$^{-2}$ pix$^{-1}$.
Interstellar extinction at 1.644 $\mu$m will be higher than at
2.122 $\mu$m (A$_{[FeII]}$/A$_{H{_2}}$ $\sim$1.5).  Also, for weak shocks, the
line flux at [FeII] 1.644 $\mu$m is typically lower than that at
H$_2$ 2.122 $\mu$m (Smith 1995).  So, for most of the faint H$_2$ features,
their non-detection in [FeII] can be attributed to the extinction and to the
shocks being relatively weak.
The lack of [FeII] emission may also be a sign of weak molecular shocks
being more common than strong atomic shocks.  H$_2$ and [FeII] line
emission are often observed in jets and bow shocks from low mass young
stars (e.g. Reipurth et al. 2000; Davis et al. 2003), although the
emission usually derives from different regions in each outflow.  The
H$_2$ is usually excited in the oblique wings of bow shocks, where low
shock velocities, low ion fractions and modest magnetic field strengths
generate bright molecular line emission.  The [FeII] is instead excited in
the compact apices of bow shocks, or in jet Mach disks, where shock
velocities are much higher and excitation conditions are more extreme.

So, although the sensitivity of our [FeII] observations is limited, and
extinction will play a role, it is nevertheless interesting that these
extreme excitation conditions (the Fe$^+$ ionization fraction is predicted
to peak at temperatures of $\sim$14,000 K [Hamann 1994], and the
critical density for collisional excitation of the 1.644\,$\mu$m [FeII]
line is $3 \times 10^4$\,cm$^{-3}$) do not seem to be prevalent across the
field observed except for B5, which is the only H$_2$ features showing
[FeII] emission.

\begin{figure}
\vspace{140pt}
\includegraphics{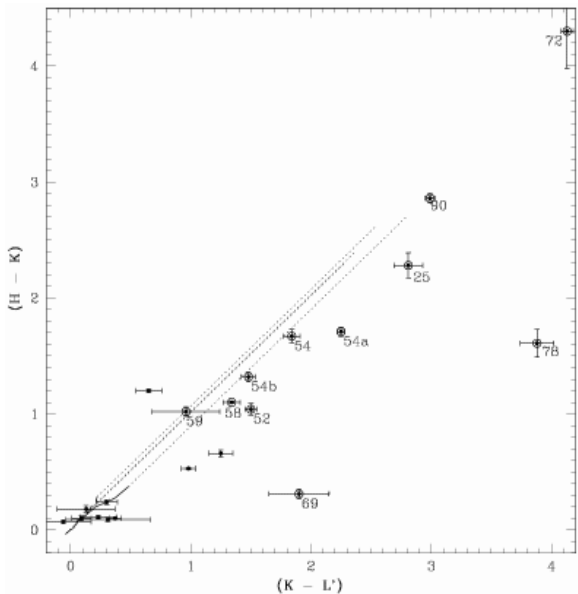}
\vspace*{105pt}
\caption{$(H-K)-(K-L')$ colour-colour diagram.  The continuous line show the
locus of main sequence stars and the broken line, that of giants.  The dotted
lines show the reddening vectors up to A$_V$=30. }
\label{g173HKKL}
\end{figure}

\vskip 5mm
\subsection{Discussion of individual sources}

\subsubsection{Outflow sources in the central cluster}

At the current spatial resolution, the central source IRS1 (\#52) itself 
is resolved to have three components, with 2 stars of similar brightness 
(\#52a and b) at a separation of 
0.40$\arcsec$$\pm$ 0.01, with the apparent line of centres oriented 
153.5$^\circ \pm$ 1$^\circ$ east of north and a much fainter component (\#52c) 
at a separation of 1.35$\arcsec$$\pm$ 0.01 from the apparent line of centres 
of the binary, at an angle 54.7$^\circ \pm$ 1$^\circ$ east of north. Contrary
to what was noticed by SW02 for IRS1, \#52a and b show IR excess in the $JHK$ 
colour-colour diagram (Fig. \ref{g173JHHK}). Both these objects are well 
detected and are the brightest ones in the $L'$ and $M'$ bands. \#52c appears
in the main sequence reddening band in the $JHK$ colour-colour diagram. It is 
located at the position of a reddened O type star.  Nevertheless, this star 
is the faintest of 
the three, being fainter than the binary components by 3.4 magnitudes in $K$.
So, if this is a reddened O star, it would be brighter in the $L'$ and
the $M'$ bands and would have been detected in our $L'$ images. The fact that 
it is not detected at wavelengths higher than $K$ implies that \#52c is not a 
reddened O star, but a star of much later type, with some amount of IR excess,
which moves it to the region of reddened O stars in Fig. {\ref{g173JHHK}}.
The position of the star in the colour-magnitude diagram (Fig. {\ref{g173KHK}})
justifies this.  We conclude that \#52c is a reddened late type star with
IR excess; the upper limit of its spectral type is $\sim$F5.  The combined
magnitude of the sources in \#52 does not show any variability.

\begin{figure*}
\vspace{165pt}
\includegraphics{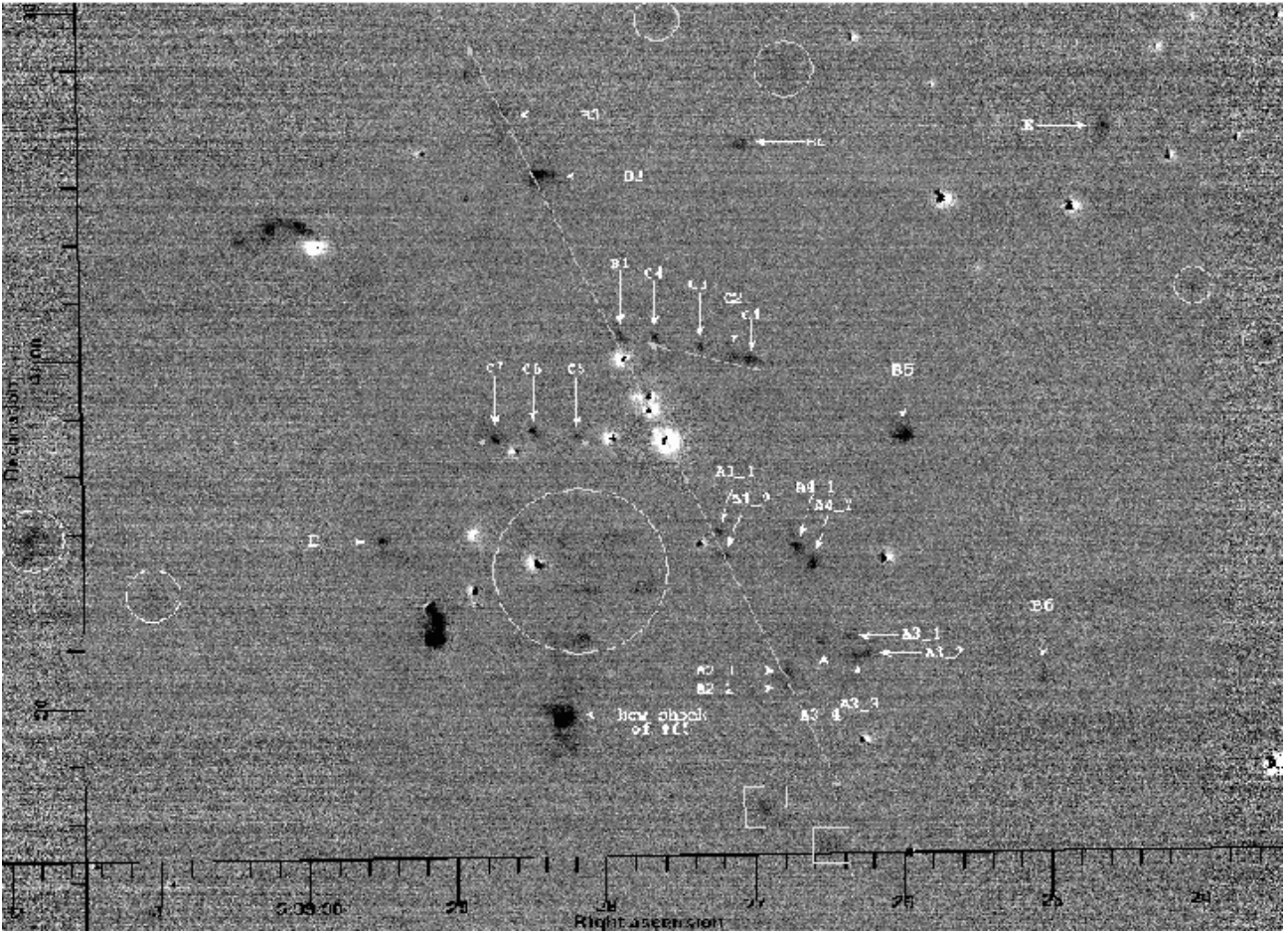}
\vspace*{230pt}
\caption{The continuum subtracted H$_2$ image of the central region. The image is shown 
as negative to enhance the H$_2$ emission features.  The 'Herbig-Haro' type 
features and bow shocks detected are labelled. Arrows are drawn along the 
aligned H$_2$ knots. The well aligned H$_2$ features strongly suggest 
collimated jets.}
\label{g17358h2}
\end{figure*}

The feature towards the north-east of \#52 (source \#19 of 
Chakraborty et al. 2002 and SW02) is not a jet, as previously noticed by SW02.
In our $JHK$ images, it is resolved into four stars,  \#54a-d, ranging from 
12.67 mag. to 17.72 mag. in the $K$ band. All four stars show IR excess. 
\#54c, the star with the least excess of the four is not detected in $L'$ and 
$M'$, whereas \#54d, the faintest of the four is detected only in the $K$ and 
$L'$ bands and is heavily reddened. \#54a-c exhibit photometric variability.
We find evidence of at least one 
(and possibly two) YSOs associated with sources \#52 and \#54.  A 
well collimated outflow extends in the north-east(NE) - south-west(SW) 
direction, which can be traced along the H$_2$ emission features labelled
B1-B3 located NE and A1-A2 located SW of the central cluster in 
Fig. {\ref{g17358h2}}.  All these features appear well detected in the $K$ 
and H$_2$ images.  The brightest of the three in the NE direction, B2, is 
marginally detected in the $H$ and $J$ bands, which are likely to be due to
H$_2$ emission lines in these bands.  B2 is much brighter in our $K$ band 
and H$_2$ images and would, if plotted in a $(J-H) - (H-K)$ colour-colour 
diagram, appear in the lower right region of the plot, as predicted by 
molecular shock models (Smith 1995).  The direction of the bow shocks B1-B3 
show that these are driven by one or more of the objects in the central 
region enclosed in the box in Fig. {\ref{g17358JHK}}, most likely from the 
binary \#52a and b. We do not see any bow shocks in the opposite direction, 
although we see sets of H$_2$ knots extending SW from source \#52, which we 
label A1 and A2, both resolved into two components each, named A1\_1, A1\_2 
and  A2\_1,A2\_2 respectively in Fig. {\ref{g17358h2}}.  These are probably 
produced by the counter jet. B1, B2 and B3 and A1 and A2 are certainly well 
aligned with \#52 and are oriented at an angle 26$^\circ$ east of north.  An 
arrow is drawn connecting them in Fig. {\ref{g17358h2}}.  The jet appears 
very well collimated with a collimation factor of at least 10, measured 
between B2 and source \#52.  Note, however, that the direction of this jet 
is different from the direction of the main outflow observed in the CO line 
emission by SW02, which was at 79$^\circ$.  The CO maps have poorer resolution 
and so, may not be resolving individual outflows.

Two other sets of H$_2$ emission features C1 - C4 and C5 - C7 are evident in
Fig. {\ref{g17358h2}}.  These are aligned in the opposite directions of the
central cluster, though they are clearly not the two lobes of the same outflow.
Two arrows are drawn in the figure in the directions of these features.
They are likely to be caused by the outflows from more than one source in \#54
and in the central cluster. Also obvious are the features marked A3 and A4 in 
the figure; A3 is resolved into two four components labelled A3\_1,A3\_2,
A3\_3,A3\_4 and A4 is resolved into two, labelled A4\_1,A4\_2.  The positions
and H$_2$ fluxes of all these features are listed in Table \ref{hhflux}}.
Together, these observations prove that the central region of 
G173.58+2.45 is indeed active in star formation and that there are multiple
objects driving outflows.

\subsubsection {Is the binary \#52a,b responsible for the outflow causing the bow-shocks B1 - B3 ?}

If the components of the visual binary \#52a,b are dynamically related, assuming
the observed separation of the binary to be a lower estimate of the major axis 
of the orbit and adopting the masses of the components to be $\sim$7.6 M$\odot$
each for the two stars for their upper limit of spectral types, B3, using 
Kepler's equation, we get a lower estimate of 1700 years for the period of the 
binary orbit.  Since the major axis could be longer and the binary period 
is more sensitive to the semi-major axis than the masses of the components,  
this period is only a lower estimate.  B1 and A1\_1, A1\_2 are at an average
distance of 10$\arcsec$ and B2 and A2\_1, A2\_2  are at an average distance 
of 24$\arcsec$ from \#52.  Assuming a jet velocity of $\sim$100 km s$^{-1}$ 
(Reipurth \& Bally 2001) in the sky plane, we get a dynamical time of 850 years 
for B1 and A1\_1, A1\_2, if they originate from \#52. The age will be higher 
than this, depending on the angle of inclination.  Similarly, B2 and A2\_1, 
A2\_2 will have a dynamic age of $\sim$2075 years or more, for an outflow 
velocity of $\sim$100 km s$^{-1}$, if they originate from \#52. This similarity
of the orbital period with the dynamical age of the outflow estimated
from the H$_2$ knots suggests 
that they might be related. All these features are probably
associated with the binary \#52a and b. Both the components must be having 
accretion disks around them. The mass accretion and outflow must be episodic,
probably occurring or getting enhanced during the periastron passage of 
the binary stars, and thereby giving rise to the features B1, B2, B3 and A1s 
and A2s. The H$_2$ image (Fig. {\ref{g17358h2}}) also shows features A3\_1-4 
and A4\_1-2, and the two features further south, which are enclosed
in boxes. It is possible that these features are related to A1 and A2
and might be caused by the precession of the jet.  But we cannot rule out
the possibility that A3s and A4s are driven by other sources in the
region, like those in \#54.  There are many other fainter H$_2$ emission 
features enclosed in circles in Fig. {\ref{g17358h2}}.  Specially noticeable 
among those are the ones towards the left half, running almost horizontally.
These features are roughly in the direction of the CO outflow (Fig. 5 of SW02).

\begin{table}
\caption{Observed fluxes of 'Herbig-Haro' type objects at the wavelength of 2.122 $\mu$m H$_2$(1-0)S(1) line in the field}
\label{hhflux}
\begin{tabular}{lll}
\hline
Feature	   &Coordinates		  &2.122 $\mu$m line flux  \\
           &RA,Dec(J2000)	  &($\times$ 10$^{-19}$ W/m$^2$) \\
\hline
B1         &5:39:27.31 35:41:01.7 &16.5           \\
B2         &5:39:27.84 35:41:15.9 &75           \\
B3         &5:39:28.06 35:41:20.3 &38           \\
B4         &5:39:26.45 35:41:18.5 &18           \\
B5         &5:39:25.38 35:40:53.0 &68           \\
B6	   &5:39:24.47 35:40:32.2 &9.6            \\
A1\_1      &5:39:26.65 35:40:45.0 &9.2            \\
A1\_2      &5:39:26.61 35:40:42.8 &17.8           \\
A2\_1      &5:39:26.21 35:40:32.9 &17.8           \\
A2\_2      &5:39:26.20 35:40:31.4 &13           \\
A3\_1      &5:39:25.78 35:40:35.8 &8.8            \\
A3\_2      &5:39:25.66 35:40:34.4 &12.6           \\
A3\_3      &5:39:25.75 35:40:34.1 &11.7           \\
A3\_4      &5:39:25.98 35:40:35.3 &10.7           \\
A4\_1      &5:39:26.12 35:40:43.8 &30.4           \\
A4\_2      &5:39:26.01 35:40:42.2 &29           \\
C1         &5:39:26.42 35:40:59.8 &19.7           \\
C2         &5:39:26.52 35:41:00.0 &15.3           \\
C3         &5:39:26.75 35:41:00.9 &10.7           \\
C4         &5:39:27.06 35:41:01.7 &11.5           \\
C5         &5:39:27.59 35:40:53.4 &7.3            \\
C6         &5:39:27.89 35:40:53.8 &13.2           \\
C7         &5:39:28.14 35:40:53.1 &19           \\
D          &5:39:28.91 35:40:44.5 &19           \\
\#65's bow shock &5:39:27.71 35:40:29.0 &205$^1$      \\
\#76       &5:39:28.57 35:40:35.9 &199$^2$      \\
\#76+neb   &		          &356$^3$      \\
E          &5:39:24.01 35:41:19.5 &29.5          \\
\hline
\end{tabular}\\
Integrated fluxes:  $^1$ in a 20 pixel radius window enclosing the emission
nebulosity near \#65, $^2$ in a 10 pixel radius window enclosing \#76abc,
$^3$ in a 33 pixel radius window enclosing \#76abc and the emission bridge between \#76 and
\#78
\end{table}

\subsubsection{\#37 and \#19}
\#37 is a deeply embedded source. The $K$ magnitude is $\sim$ 16.5 and is
variable. Our $K$ image shows a faint 'comet' shaped nebulosity associated
with it, which disappears in the continuum-subtracted H$_2$ image. The
source is faint, but detected in the $L'$ band.

Source \#19 is a visual binary, both the components of which (\#19a and b, 
separated by 0$\arcsec$.96) appear reddened.  There is a nebulosity 
around them in the $K$ band, which disappears in the continuum-subtracted 
H$_2$ image.  The component stars are not detected in the $H$ band, but a faint
nebulosity is present. The source is bright in the $L'$ band, most of the 
contribution appearing to come from \#19a.

\subsubsection{Outflow source \#65}
\#65 is detected only in the $K$ band and is a deeply embedded.
There is a strong 'cap' like emission in H$_2$ towards the north-east of 
the source at a separation of $\sim1\arcsec$. This feature 
appears split into two and the continuum subtracted H$_2$ image shows that it 
is mostly emission in H$_2$. It is very likely that \#65 drives an outflow, 
which is highly inclined with respect to the sky plane. The 'cap like' feature 
in H$_2$ is probably the bow shock of the blue-shifted lobe of the outflow, 
inclined in our direction.  Extinction probably obscures our view of the 
red-shifted lobe.  There are two other extremely red sources, \#60 and \#62, 
situated NW of \#65, which are well detected in $K$ and are 
brighter than \#65.  The sources appear deeply embedded and are marginally 
detected in the $L'$ band. There is faint H$_2$ emission very close to \#62
and the star exhibits variability in $K$.

\subsubsection{The extremely red object, \#72}
\#72 is one of the reddest objects in 
in the region.  This source was detected in all our $K$ band images.
It was not detected in the $J$ band and was only weakly detected in $H$
on 2002, Oct. 22, when the seeing was $<$0.4 and the integration was deep.
It is  bright in the $L'$ and $M'$ bands and it stands
out in both the $(H-K) - K$ colour-magnitude diagram (Fig. \ref{g173KHK}) and 
in the $(H-K) - (K-L')$ colour-colour diagram (Fig. \ref{g173HKKL}).  The 
locations of the object in these diagrams give the appearance of a late O or
early B star with very high extinction.  Since the object was not detected
in the $H$ band on 2003 March 19, we have used the $L'$ magnitude of 2003,
March 19 along with the $HK$ magnitudes of 2002 Oct. 22 to place this star
on the $(H-K) - (K-L')$ colour-colour diagram. So, the location of the star
close to the reddening band in Fig. \ref{g173HKKL} cannot be considered as
due to a lack of IR excess.  The $H$ band magnitude has a large uncertainty
of 0.32 mag and the the object exhibited variability in $K$ during the three
epochs of our observations.  Note that $(H-K)$ 
colour also will not be free from IR excess.
This object is very likely to be a YSO with 
a considerable amount of circumstellar matter; indeed, the $K$ band image
shows a well defined 'comet' shaped nebula associated with it.
We detect no H$_2$ emission from this nebula, which is probably 
dominated by reflection and/or thermal emission by dust. The source is as 
close to the IRAS positions as \#52.  

\subsubsection{Outflows from the embedded binary, \#78}
This is a set of interesting objects, with \#76 located south-west
of \#78. The region is shown in detail in Fig. {\ref{g17358region3}}. Both 
sources are resolved into three components each, with \#78 showing a tighter 
association.  The components of \#78 are named \#78a,b and c based on their 
decreasing brightness in the $K$ band.  \#78a,b appear to be point sources, but 
\#78c is slightly extended and may be part of an outflow. The components of
\#76 are labelled \#76a,b and c.  \#76c disappears in the continuum 
subtracted H$_2$ image (Fig. \ref{g17358region3}) while \#76a and b appear like
point sources.  There is a bridge of nebulosity connecting \#76 and 
\#78, which is mostly H$_2$ emission.  The projected separation between \#76 
and \#78 is 3$\arcsec$, which is 0.026 pc (5400 AU) at a distance of  1.8 kpc.

Though both \#76 and \#78 are of comparable brightness in $K$, their $JHK$ 
colours (Fig. \ref{g173JHHK}) show that \#76 is more reddened than \#78. 
There is no trace of \#76 in the $L'$ and $M'$ images.  \#78 is well detected 
in both $L'$ and $M'$ bands and the IR colours place it in the region of the 
colour-colour diagrams (Figs. \ref{g173JHHK}, \ref{g173HKKL}) occupied by YSOs. 
So, \#78 is probably a set of two young outflow sources, which are responsible 
for the H$_2$ emission in Fig. \ref{g17358region3} (around \#76 and between 
\#76 and \#78). Note that there is a faint emission (D) in the counter-flow 
\begin{figure}
\vspace{140pt}
\includegraphics{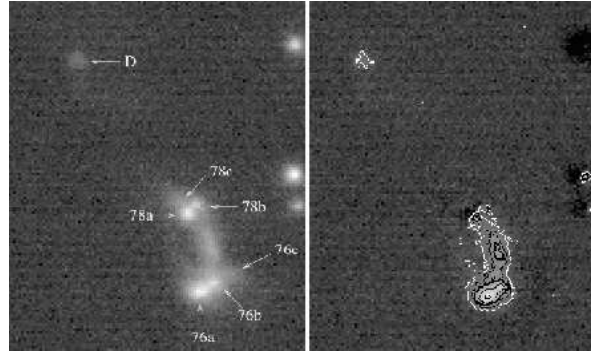}
\vspace*{10pt}
\caption{The left portion of the figure shows the observed H$_2$ image
and the right portion shows the continuum subtracted H$_2$ image of
\#76, \#78 and the emission in between. The contours are at flux levels
3.0, 1.5, 0.8, 0.5 and 0.25 $\times$10$^{-19}$ Wm$^{-2}$pix$^{-1}$
(1 pixel of UFTI=0.00825 arcsec$^2$).}
\label{g17358region3}
\end{figure}
direction. Both \#76 and \#78 appeared to be
continuously fading in the $JHK$ bands within the period of our observations,
with \#76 fading faster than \#78. \#78 exhibited fading in the $L'M'$ bands 
too.  The coordinated fading in the $JHK$ bands implies that \#78 and \#76 are
probably related and the emission from \#76 and and the bridge between \#76 
and \#78 depend on \#78 for the source of energy.  However, 
Table {\ref{JHKPhot}} shows 
that on 2000 Dec. 26, \#76 was brighter than \#78 in the $J$ and $K$ bands.  
This scenario is possible if the emission from \#76 is mostly contributed
by line emission from shocked H$_2$.  (The fact that \#76c disappears in the 
H$_2$ image upon continuum subtraction shows that there is also some amount
of reflection and/or emission from dust contributing to \#76). It is also 
likely that \#76-\#78 direction is inclined very much with respect to the 
sky plane so that \#76 suffers from much less extinction 
compared to \#78 in an inhomogeneous medium.  The much fainter H$_2$ 
emission in the opposite direction strengthens this argument.  Alternatively,
there is a possibility that \#78 and \#76 are two different sets of objects, 
the medium around \#76 emits mostly in H$_2$ and that the coordinated 
variability that we see is merely a coincidence.  But, considering the fact
that \#76 is not detected in the $L'$and $M'$ bands even though it is nearly
as bright as \#78 in the $JHK$ bands and that it lies in the middle-right
portion of the $(J-H)-(H-K)$ colour-colour diagram, close 
to the region occupied by reddened shocks (Smith 1995),
it is more likely to be composed of reddened bow shocks of the outflow 
from \#78.

From their spatial locations and $HKL'M'$ colours, it seems
likely that \#72 and \#78 contribute significantly to the IRAS fluxes.
These two sources could be two of the prominent candidates for 
driving the CO outflow seen in the field by SW02.

\subsubsection{The variable outflow source, \#90}

In Fig. \ref{region2h2}, we focus on the region around \#90.
Chakraborty et al. (2000) first detected the variability of source \#90
(their \#12). They proposed that this might be due to FU Orionis type behavior. 
Later NIR observations by SW02 confirmed the variable nature of this star. 
Our multi-epoch NIR observations also confirm the variability of this
star. All the available $JHK$ magnitudes of this object are compiled in 
Table {\ref{JHKall}}, which shows the variability of the source at these 
wavelengths. The $J$ and $H$ band filters used by all the investigators are
similar.  Chakraborty et al. (2000) observed using a $K'$ filter and  SW02
and 2MASS used $K_{s}$ filter which both have their central wavelengths
shorter than 2.2 $\mu$m of the $K$ filter used by us.
An inspection of the magnitudes estimated for
some of the bright non-variable stars in the field shows that our estimates
agree with that of 2MASS within 0.04 mag in all bands, with that of Chakraborty et al.
within 0.1 magnitudes in the $J$ band and within $\sim$0.12 mag in $H$ and $K$ and
with SW02 within 0.05 mag in the $J$ band and $\sim$0.1 mag in the $H$ and $K$ bands. 
The differences between the photometric systems is
less than these and are ignored. The photometric variability that we see on this
source is higher than these differences and the photoemtric errors.
It can be seen from Table {\ref{LMPhot}} that the object
shows co-ordinated variability in the $L'$ and $M'$ bands also over the
two epochs of our observations.  In Fig. {\ref{source90var}}, we have plotted 
the observed $JHK$ magnitudes of this star against the Julian dates. The 
calendar days are also shown in this diagram.  Magnitudes quoted by
Chakraborty et al. (2000), SW02 and from 2MASS are included along with ours.
Together, these give 6 epochs of data in $K$, from 1997 October to 2003 March.  
The object is detected in $J$ only on those epochs when it is brightest in $K$.
It is too early to say if there is any periodicity in its variability.
The nature of the variability is unlike that of a typical FU Ori type, which
show longer time scales, of the order of tens of years.  (A review of 
the properties of FU Ori objects can be seen in Hartmann \& Kenyon 1996).

This star has large $JHK$ colours and large IR excess, as can
be seen in Figs. {\ref{g173JHHK}} and {\ref{g173HKKL}}. It also shows an
extended, well collimated, nebulosity, which appears to originate from
the star.  The nebulosity is not seen in the $H$ band and is seen well in
the $K$ band and H$_2$. The continuum subtracted H$_2$ image shows that the
emission in the nebulosity is mostly at 2.122 $\mu$m.
This implies that the emission is mostly from shocked molecular Hydrogen.
The large infrared colours and the nature of the nebulosity tells that this
object is a YSO, with an accretion disk and an outflow.

\begin{figure}
\vspace{160pt}
\includegraphics{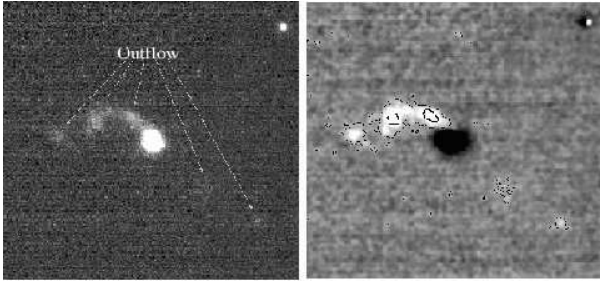}
\vspace*{-34pt}
\caption{The left portion of the figure shows the observed H$_2$ image
of star $\#$90 and its outflow. The right portion shows the continuum 
subtracted, smoothed, H$_2$ image on which line intensities are plotted 
as contours.
The contours are at 2.0 and 0.5 $\times$ 10$^{-20}$ Wm$^{-2}$pix$^{-1}$ 
respectively. }
\label{region2h2}
\end{figure}

\begin{table}
\caption{Observed magnitudes of Source \#90}
\label{JHKall}
\begin{minipage}{140mm}
\begin{tabular}{lllll}
\hline
Cal. date           &JD           &$J$            &$H$            &$K$              \\
                        &2400000+       &&&\\
\hline 
97Oct18$^1$         &50739.5      &17.99(.04)     &14.70(.01)     &12.35(.01)     \\
98Feb03$^2$         &50847.5      &17.15(.18)     &14.29(.04)     &12.01(.01)     \\
00Jan10$^3$         &51553.5      &               &15.32(.07)     &13.14(.07)     \\
01Dec25$^4$         &52270.01     &               &               &14.01(.02)     \\
02Oct22$^4$         &52570.06     &               &15.16(.02)     &12.26(.01)     \\
02Nov06$^4$         &52584.98     &18.73(.04)     &               &               \\
03Mar19$^4$         &52717.8      &               &16.09(.03)     &13.23(.02)     \\
\hline 
\end{tabular}
\end{minipage}\\
$^1$SW02, $^2$2MASS,  $^3$Chakraborty et al. (2000), $^4$present work
\end{table}

Left portion of Fig. {\ref{region2h2}} shows the observed H$_2$ image
of \#90. The brighter part of the outflow points towards the NE and has a
twisted appearance;  careful inspection shows that there is also some H$_2$ 
emission in the SW direction of this star, though this is much fainter
than the NE component.  We are probably seeing a bipolar outflow,  which is
somewhat inclined with respect to the line-of-sight. The differential
extinction due to the inclination and the inhomogeneity of the  surrounding
medium must be contributing to the large difference in brightness between the 
NE and SW outflows. The continuum subtracted H$_2$ image of the region is 
Gaussian smoothed with 
\begin{figure}
\includegraphics{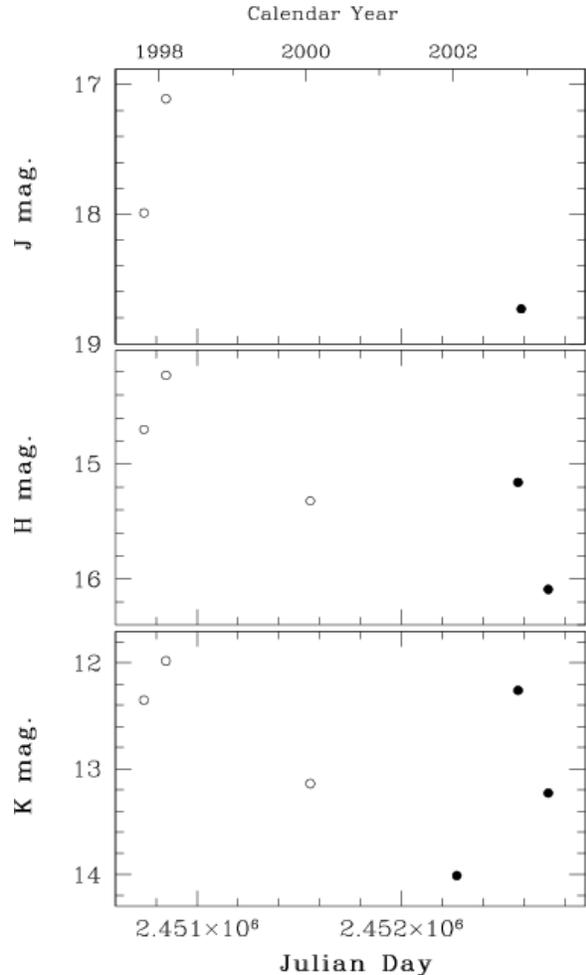}
\vspace*{375pt}
\caption{Variation in the $JHK$ magnitudes of star \#90.  The open circles
show the previous observations and the filled circles show our observations.}
\label{source90var}
\end{figure}
a Gaussian of 3 pix FWHM to enhance the contrast of the faint H$_2$ emission,
especially the SW lobe, in the presence of the background noise.
The right half of Fig. {\ref{region2h2}}
shows the continuum subtracted, smoothed H$_2$ image on 
which line fluxes are plotted as contours.  It is interesting 
to note that the direction of the brighter lobe of the outflow observed in the
H$_2$ image matches roughly with one of the blue-shifted emission lobes 
(in the NE direction) in the CO line emission map of SW02.  The twisted nature 
of the NE outflow suggests that there is probably a precession of the 
object or jet, which results in time dependent variations in the direction of
the outflow. Alternatively, we are looking at a partially shocked shell at
the edges of a parabolic outflow. High resolution spectroscopy of the outflow 
at 2.122 $\mu$m would shed further light on the geometry of this system.

The $JHK$ light curves (Fig. {\ref{source90var}}) show that the flux 
variations are coordinated in all three bands.  The first pair of points 
(97Oct18, 98Feb03) show the variations increasing towards shorter wavelengths. 
This type of colour dependence of the variability in the $JHK$ bands can be 
caused by obscuration by dust clouds passing in front of the star, which 
spiral in towards the star when accreted.  However, if the variations are 
purely due to variable extinction, then we should see a shift
in the position of the star parallel to the reddening vectors in the 
$(H-K) - K$  colour-magnitude diagram in Fig. {\ref{g173KHK}} when we plot 
the data from the different epochs. We have plotted all the available 
simultaneous $HK$ observations of the source from Table {\ref{JHKall}}
in the figure. 90$_1$ and 90$_2$ show the location of our observations;
90$_3$ shows the observation of Chakraborty et al. (2000); 90$_4$, the 2MASS 
observations and 90$_5$, that of SW02.  We can see from the figure that,
though 90$_1$, 90$_4$ and 90$_5$ are located close to a reddening line,
90$_2$ and 90$_3$ deviate very much.  Accordingly, the pairs 
(98Feb03, 00Jan10) and (02Oct22, 03Mar19) do not show any colour dependence 
in the $H$ and $K$ magnitude variations in Fig. {\ref{source90var}}, though 
though they vary simultaneously.  Thus, the shift in the 
location of the object in the colour-magnitude diagram is not
caused by reddening alone.  Instead, the variability is likely to be
due to a combination of mechanisms, like changes in the accretion luminosity
due to variable accretion and variable reddening caused by large clumps of
dusty material crossing the line of sight when they spiral in towards the 
central star.  At this stage, we cannot say if there is any binarity involved.
The large reddening and the observed bipolar H$_2$ emission imply that the
object is in an actively accreting phase.  In these type of objects, it is 
natural that the accretion luminosity may be comparable to the stellar 
luminosity (Herbst \& Shevchenko 1999).  In the $(J-H) - (H-K)$ colour-colour
diagram, the object is located towards the region occupied by luminous YSOs 
(Lada \& Adams 1992). Observations at longer wavelengths are required for 
a proper estimate of the luminosity of this object.  More observations of 
this star are being planned.

\subsection{Observations at longer wavelengths}

The region was imaged in the IRAS mission. The IRAS positions 
(IRAS 05361+3539, $\alpha$=05:39:27.7, $\delta$=+35:40:43, J2000) are east 
of \#52 by 10$\arcsec$.5 and south by 10$\arcsec$.3.  Source \#72 
(with the cometary nebula) is the object closest to the IRAS positions; 
\#72 also has strong
colours in the $JHKL'M'$ bands.  It is well detected only in $K$ or above; 
in $H$, it is only marginally detected. The IRAS position is marked by a '+' 
sign in Fig. {\ref{g17358JHK}}. The error ellipse of the IRAS positions is
(32, 7, 86), ie., the source can be said to be located within an ellipse
of semi-major axis 32$\arcsec$ and semi-minor axis 7$\arcsec$ with the 
semi-major axis inclined at an angle of 86$^{o}$ E of North, with 95\% 
confidence. Looking at the $L'$ and $M'$ images observed by us and their 
location with respect to the IRAS coordinates, it is highly likely that 
the flux measured by IRAS is the combined emission by a number of stars, 
particularly \#72, \#78, \#52 and \#54.

SW02 detected two mm wavelength continuum emission sources, MM1 and MM2,
using interferometric observations at 2.7 mm within a field of 
5$\arcmin$.3 $\times$ 3$\arcmin$,  which were interpreted as due to thermal
dust emission associated with embedded protostars. The location of MM1
($\alpha$=05:39:28.67, $\delta$=+35:40:37.8) matches closely with \#78 in 
our near IR images, 
located between \#78 and \#76.  From the fact that \#76 is not detected in 
the $L'$ and $M'$ bands,  \#78 is the most likely counterpart of MM1.  The
location of MM2 ($\alpha$=05:39:27.08, $\delta$=+35:40:56.1) matches well 
with \#54, being located between \#54 a,b \& d.  All four objects in \#54 come 
within their beam.  We cannot say for sure if there is no contribution to MM2
from  \#52. Assuming optically thin dust emission, they determined a mass of
12 M$_\odot$ for the gas and dust associated with MM1 and 7M$_\odot$  for
MM2.  This is consistent with the larger $JHK$ colours observed for \#78 
compared to the colours of the components of \#54.  Also the larger amount 
of nebulosity and the indications of outflow from H$_2$ emission 
implies that \#78 is younger than \#54 and has more gas and dust 
associated with it.

Methyl Cyanide (CH$_3$CN) emission is observed from warm, dense molecular 
cloud cores, which are believed to be heated by embedded massive protostars
during the pre-UCHII phase of massive star formation. The 220 GHz CH$_3$CN 
survey of Pankonin et al. (2001) failed to detect any emission from the 
vicinity of IRAS 05361+3539, showing that either the emission is too faint or
the IRAS source or the objects near it have evolved past the very early stages.
In a CH$_3$OH maser survey conducted by Blaszkiewicz \& Kus (2004), 12.2 GHz
maser emission was found from this region with a peak flux density of 12.5 Jy
at a position slightly different from the position of the 6.7 GHz emission 
observed by Menten (1991). They did not detect any 6.7 GHz emission, though 
Menten (1991)  detected a weak 6.7 GHz maser source (peak 5Jy) in this region.  
The central coordinates of Blaszkiewicz \& Kus (2004) match closely with the 
centre of our IR images, whereas that given by Menten (1991) are south of 
this position by $\sim$ 10$\arcmin$ and may not coincide with that of 
Blaszkiewicz \& Kus (2004).  Note that Menten's observations used a 
beam size of 5$\arcmin$ whereas the observations by Blaszkiewicz \& Kus 
have a beam size of 3$\arcmin$ at 12.2 GHz and 5$\arcmin$.5 at 6.7 GHz.
Also the line-of-sight radial velocities of both detections are
different.  12.2 GHz maser emission is associated with class II 6.7 GHz
masers,  which are usually observed towards UCHIIs, whereas class I 6.7 GHz 
masers are associated with outflow shocks (Blaszkiewicz \& Kus 2004,  
Plambeck \& Menten 1990). This implies that there is a UCHII region present 
towards the centre of the field, though we do not identify this object 
specifically in our IR images.

\section{Conclusions}

Detailed multi-epoch photometry of the intermediate-mass YSO, G173.58+2.45,
and its surrounding cluster is performed at high spatial resolution in 
the infrared $JHKL'M'$ photometric bands and at the wavelengths of 
2.122 $\mu$m H$_2$(1-0) S(1) and 1.644 $\mu$m [FeII] using narrow band 
filters. Through the infrared colours and the presence of bow shocks and other 
outflow features seen in the H$_2$ images, we identify many of the outflow 
sources.  It is shown that the outflow seen in this region in low-resolution
CO maps is due to multiple collimated outflows.

The source detected by IRAS (IRAS 05361+3539) is probably the combined 
emission from a number of YSOs near the central part of the field; the 
prominent ones appear to be the sources \#72, \#78ab, \#52ab and \#54abd.

From the apparent separation of the binary components \#52ab and an assumption
about their masses, an estimate of the lower limit of the 
orbital period is made.  If we assume a typical flow velocity of 
100 km s$^{-1}$, there is a reasonably good agreement between the binary 
period and the difference in 
the dynamical times of the individual sets of aligned H$_2$ emission 
features. This shows that the mass accretion and the associated outflow are
episodic and must be occurring or getting enhanced during the periastron 
passage of the binary system.  However, this source doesn't appear to be the 
the main contributer to the outflow discovered in the CO emission lines.

Source \#90 is associated with a well defined outflow, both the lobes of 
which are detected in H$_2$ emission. The presence of these outflow lobes
indicate that there is ongoing accretion.  This object shows large 
photometric variability.  We do not have a sufficient number of 
observations of this star to look for periodicity.  However, the 
very twisted appearance of the outflow may be an indication of the precession
of the object or the jet or of the presence of a partially shocked shell at 
the edges of a parabolic outflow.  The large photometric variability
could be due to varying amount of accretion.  An obscuration mechanism alone 
is not sufficient to explain the nature of the photometric variability. 
Instead, the variability could be due to a 
combination of phenomena, like variations in the accretion luminosity and 
variable obscuration from accreting circumstellar matter.  From the present 
data, we cannot establish whether this object has a companion. The object 
definitely needs follow-up photometric observations at shorter time intervals 
and at high angular resolution for a more complete understanding.

Two objects (\#72 and \#90) exhibit IR colours 
similar to that of massive YSOs. Observations at longer wavelengths at high
spatial resolution are required to estimate their luminosities.

Several other YSOs are detected in the field from the H$_2$ images and 
the IR colours.  Many of these YSOs show photometric variability. 
The YSO candidates detected appear to be in different stages of their 
protostellar evolution.  These observations show that this is a region 
very active in star formation and that the star formation is not coeval.

\section*{Acknowledgments}
UKIRT is operated by the Joint Astronomy Centre, Hilo, Hawaii, on behalf 
of the U.K. Particle Physics and Astronomy Research Council.  We would
like to thank the UKIRT Service Observing Programme for obtaining some of 
the data. Several of the software packages developed by the  
Starlink project run by CCLRC on behalf of PPARC and IRAF developed by
NOAO are used for reducing the data.  We have made use of 2MASS data obtained 
as a part of Two Micron All Sky Survey, a joint project of University of 
Massachusetts and the Infrared Processing and Analysis Centre/California 
Institute of Technology.  This research has also made use of SIMBAD database
operated by CDS, Strasbourg, France. We thank the referee
for many valuable comments,  which have improved the quality of
the paper.

\appendix

\section[]{$JHK$ Photometric observations}
\begin{table*}
{\small
\begin{minipage}{500mm}
\caption{$JHK$ Photometric magnitudes of sources within 3$\arcmin$$\times$3$\arcmin$ of IRS1.  The error estimates are shown against the estimated magnitudes.}
\label{JHKPhot}
\begin{tabular}{@{}llllllllllll}
\hline
No:   &SW02  &RA &Dec &\multicolumn{3}{c}{---------------- $J$ Band ---------------- }&\multicolumn{2}{c}{-------- $H$ Band --------}&\multicolumn{3}{c}{---------------- $K$ Band ----------------} \\
      &{ID\footnote{Source identifications from Chakraborty et al. (2000) and SW02}}    &J2000      &J2000              &20011226   &20021106   &20030319    &20021022 &20030319   &20011226 &20021022 &20030319       \\
\hline
1     &      &5:39:20.73 &35:41:27.8    &         &         &          &         &16.33 .06&         &         &15.50 .08  \\[-.1mm]
2     &      &5:39:20.85 &35:41:04.3    &         &         &17.39  .14&         &16.87 .07&         &         &16.32 .12  \\[-.1mm]
3     &26    &5:39:21.60 &35:40:21.9    &         &         &          &         &16.15 .06&         &         &15.41 .08  \\[-.1mm]
4     &      &5:39:21.98 &35:41:13.4    &         &         &          &19.49 .15&         &         &18.09 .05&           \\[-.1mm]
5     &      &5:39:22.09 &35:41:47.3    &         &         &          &         &         &         &17.68 .05&           \\[-.1mm]
6     &16    &5:39:22.36 &35:40:57.4    &         &16.70 .05&17.00  .12&16.22 .03&16.42 .06&         &15.84 .02&16.05 .10  \\[-.1mm]
7     &24    &5:39:22.39 &35:40:34.8    &         &16.52 .04&16.57  .10&15.30 .02&15.28 .02&         &14.60 .01&14.60 .02  \\[-.1mm]
8     &      &5:39:22.54 &35:41:10.7    &         &17.97 .12&          &17.52 .06&         &         &17.13 .03&           \\[-.1mm]
9     &      &5:39:22.56 &35:40:06.3    &         &         &          &         &         &         &18.67 .12&           \\[-.1mm]
10    &      &5:39:22.60 &35:40:56.6    &         &         &          &18.55 .05&         &         &16.90 .05&16.95 .12  \\[-.1mm]
11    &      &5:39:22.68 &35:40:20.8    &         &         &          &19.72 .20&         &         &18.76 .11&           \\[-.1mm]
12    &1     &5:39:22.82 &35:41:28.6    &         &16.36 .05&16.07  .05&15.12 .02&15.13 .02&         &14.47 .01&14.46 .04  \\[-.1mm]
13    &      &5:39:22.92 &35:39:53.8    &         &         &17.67  .15&         &         &         &         &16.86 .12  \\[-.1mm]
14    &2     &5:39:22.92 &35:40:23.9    &         &         &10.37  .02&         &10.11 .02&         &         &10.01 .01  \\[-.1mm]
15    &3     &5:39:22.93 &35:41:41.8    &         &12.89 .00&12.85  .02&12.63 .01&12.61 .02&         &12.52 .01&12.52 .02  \\[-.1mm]
16a   &      &5:39:22.99 &35:41:27.9    &         &         &          &19.08 .10&         &         &17.59 .04&           \\[-.7mm]
16b   &      &5:39:23.01 &35:41:27.4    &         &         &          &         &         &         &17.91 .05&           \\[-.7mm]
{\bf 16}&&\multicolumn{2}{c}{\bf 16(a+b)+nebula}& &         &          &17.98 .10&         &         &16.46 .15&17.03 .11 \\[-.1mm]
17    &      &5:39:23.09 &35:40:04.0    &         &         &          &         &         &         &19.60 .25&           \\[-.1mm]
18    &17    &5:39:23.11 &35:41:18.3    &16.67 .07&16.50 .04&16.64  .05&15.88 .04&15.96 .03&15.64 .01&15.50 .01&15.65 .04  \\[-.1mm]
19a   &      &5:39:23.39 &35:41:28.6    &         &         &          &         &         &16.16 .05&16.20 .04&           \\[-.7mm]
19b   &      &5:39:23.33 &35:41:29.2    &         &         &          &         &         &16.62 .05&16.56 .10&           \\[-.7mm]
{\bf 19}& &\multicolumn{2}{c}{\bf 19(a+b)+nebula}&&         &          &         &         &14.55 .10&14.61 .10&14.65 .10  \\[-.1mm]
20    &      &5:39:23.37 &35:41:14.1    &         &         &          &         &         &17.37 .06&17.51 .03&           \\[-.1mm]
21    &      &5:39:23.54 &35:40:27.0    &         &         &          &20.00 .17&         &         &18.80 .09&           \\[-.1mm]
22    &      &5:39:23.55 &35:41:10.2    &         &         &          &         &         &         &18.17 .10&           \\[-.1mm]
23    &      &5:39:23.57 &35:41:16.7    &18.97 .15&18.89 .12&          &16.09 .01&16.14 .03&14.58 .01&14.57 .01&14.57 .02  \\[-.1mm]
24    &      &5:39:23.59 &35:40:50.6    &         &         &          &         &         &         &18.57 .09&           \\[-.1mm]
25    &      &5:39:23.64 &35:41:26.0    &         &         &          &17.44 .11&17.52 .11&15.17 .03&15.16 .02&15.14 .05  \\[-.1mm]
26    &      &5:39:23.73 &35:41:43.9    &         &19.65 .12&          &18.04 .04&         &         &17.02 .03&16.96 .10  \\[-.1mm]
27    &      &5:39:23.83 &35:40:00.3    &         &19.30 .12&          &18.20 .04&         &17.17 .15&17.30 .04&           \\[-.1mm]
28    &46    &5:39:23.97 &35:39:29.3    &         &         &17.50  .10&         &16.41 .06&          &         &15.89 .06  \\[-.1mm]
29    &      &5:39:23.98 &35:40:54.6    &         &         &          &         &         &         &19.10 .09&           \\[-.1mm]
30    &      &5:39:24.01 &35:41:25.2    &         &         &          &20.37 .25&         &         &18.74 .08&           \\[-.1mm]
31    &5     &5:39:24.17 &35:42:02.9    &         &         &13.02  .02&         &12.52 .02&         &         &12.28  .01 \\[-.1mm]
32    &4     &5:39:24.24 &35:41:12.4    &16.39 .05&16.44 .04&16.40  .05&14.28 .01&14.22 .03&13.06 .01&13.08 .01&13.10  .02 \\[-.1mm]
33    &      &5:39:24.34 &35:42:09.0    &         &         &          &         &17.11 .09&         &         &15.41  .04 \\[-.1mm]
34    &      &5:39:24.47 &35:41:27.3    &         &         &          &         &         &         &19.11 .20&           \\[-.1mm]
35    &      &5:39:24.57 &35:40:39.8    &19.14 .15&19.05 .12&          &18.32 .03&         &18.10 .08&18.08 .06&           \\[-.1mm]
36    &      &5:39:24.83 &35:40:01.2    &         &19.55 .20&          &         &detected &         &         &           \\[-.1mm]
37    &      &5:39:24.87 &35:41:07.2    &         &         &          &         &         &16.93 .10&16.55 .01&16.72 .06  \\[-.1mm]
38    &6     &5:39:25.09 &35:41:13.2    &12.54 .02&12.55 .02&12.56  .02&12.32 .01&12.32 .02&12.25 .02&12.25 .01&12.23 .02  \\[-.1mm]
39    &      &5:39:25.16 &35:41:23.2    &         &         &          &17.95 .06&         &16.17 .08&16.09 .06&16.22 .07  \\[-.1mm]
40    &13    &5:39:25.39 &35:40:16.8    &14.97 .02&14.96 .02&15.08  .04&14.49 .03&14.50 .02&14.30 .03&14.38 .01&14.35 .03  \\[-.1mm]
41    &7     &5:39:25.53 &35:40:42.3    &14.11 .02&14.12 .02&14.20  .02&13.76 .02&13.78 .02&13.64 .02&13.67 .01&13.66 .02  \\[-.1mm]
42    &      &5:39:25.61 &35:39:49.8    &         &         &          &19.57 .13&         &         &18.87 .10&           \\[-.1mm]
43    &      &5:39:25.63 &35:40:08.3    &         &         &          &         &         &         &19.77 .17&           \\[-.1mm]
44    &14    &5:39:25.69 &35:40:26.6    &17.44 .07&17.38 .05&17.75  .10&15.63 .02&15.75 .03&14.58 .02&14.60 .01&14.70 .05 \\[-.1mm]
45    &18    &5:39:25.69 &35:41:27.2    &16.61 .04&16.67 .04&16.75  .06&15.45 .02&15.52 .03&14.50 .02&14.54 .01&14.59 .04  \\[-.1mm]
46    &      &5:39:25.79 &35:40:30.4    &         &         &          &19.95 .11&         &         &18.51 .06&           \\[-.1mm]
47    &      &5:39:25.73 &35:42:06.9    &         &         &          &         &         &         &         &16.51 .08  \\[-.1mm]
48    &      &5:39:25.73 &35:42:07.0    &         &         &          &         &detected &         &         &16.86 .14  \\[-.1mm]
49    &      &5:39:25.94 &35:39:13.0    &         &         &          &         &         &         &         &16.51 .14  \\[-.1mm]
50    &15    &5:39:26.77 &35:40:43.8    &15.64 .03&15.63 .02&15.68  .06&15.04 .01&15.08 .02&14.91 .03&14.90 .01&14.84 .02  \\[-.1mm]
51    &      &5:39:26.52 &35:41:31.8    &         &         &          &19.63 .09&         &         &18.74 .07&           \\[-.1mm]
52a   &8     &5:39:26.98 &35:40:52.8    &         &13.80 .01&          &12.58 .01&         &11.53 .02&11.63 .01&           \\[-.8mm]
52b   &8     &5:39:26.99 &35:40:52.5    &         &14.16 .02&          &12.87 .01&         &11.75 .02&11.76 .01&           \\[-.8mm]
52c   &8     &5:39:27.08 &35:40:53.5    &         &17.21 .02&          &15.95 .02&         &14.89 .08&15.10 .02&           \\[-.8mm]
{\bf 52}&8 &\multicolumn{2}{c}{\bf(a+b+c)}&13.18 .03&13.18 .03&13.16 .03&11.88 .04&11.85 .04&10.82 .05&10.84 .03&10.85 .05  \\[-.1mm]
53    &36    &5:39:27.08 &35:39:27.4    &         &         &          &         &17.23 .12&         &         &16.16 .10  \\[-.1mm]
54a   &19    &5:39:27.09 &35:40:55.4    &16.78 .04&16.68 .03&16.72  .08&14.47 .02&14.44 .03&12.80 .02&12.67 .01&12.73 .02  \\[-.8mm]
54b   &19    &5:39:27.09 &35:40:56.6    &17.24 .04&16.89 .03&16.80  .08&15.24 .03&14.78 .03&13.89 .02&13.93 .02&13.46 .03  \\[-.8mm]
54c   &19    &5:39:27.17 &35:40:56.5    &17.84 .05&17.87 .04&18.10  .20&16.28 .05&15.97 .04&14.97 .03&15.04 .03&14.81 .05  \\[-.8mm]
54d   &19    &5:39:26.97 &35:40:56.9    &         &         &          &         &         &         &17.72 .05&           \\[-.8mm]
{\bf 54}&19&\multicolumn{2}{c}{\bf (a+b+c+d)+nebula}&15.66 .04&15.66 .03&15.64 .03&13.75 .04&13.60 .04&12.17 .05&12.08 .04&12.06 .05  \\[-.1mm]
\hline
\end{tabular}
\end{minipage}
}
\end{table*}
\begin{table*}
{\small
\begin{minipage}{500mm}
\contcaption{}
\begin{tabular}{@{}llllllllllll}
\hline
No:   &SW02  &RA &Dec &\multicolumn{3}{c}{---------------- $J$ Band ---------------- }&\multicolumn{2}{c}{-------- $H$ Band --------}&\multicolumn{3}{c}{---------------- $K$ Band ----------------} \\
      &ID    &          &               &20011226   &20021106   &20030319    &20021022 &20030319   &20011226 &20021022 &20030319       \\
\hline
55    &      &5:39:27.13 &35:41:28.8    &18.69 .08&18.78 .08&          &18.08 .10&         &17.80 .06&17.84 .03&           \\[-.1mm]
56    &      &5:39:27.16 &35:42:07.8    &         &         &17.15  .10&         &16.64 .06&         &         &16.21 .08  \\[-.1mm]
57    &      &5:39:27.18 &35:40:40.6    &18.50 .08&18.47 .02&          &17.80 .02&         &17.53 .04&17.46 .02&           \\[-.1mm]
58    &9     &5:39:27.27 &35:40:59.8    &15.35 .03&15.38 .02&15.34  .05&13.89 .01&13.90 .03&12.81 .01&12.86 .01&12.80 .02  \\[-.1mm]
59    &20    &5:39:27.36 &35:40:53.1    &16.18 .03&16.20 .01&16.37  .08&14.50 .01&14.52 .03&13.52 .03&13.41 .01&13.50 .03  \\[-.1mm]
60    &      &5:39:27.38 &35:40:32.4    &         &         &          &         &         &17.00 .20&17.10 .10&           \\[-.1mm]
61    &      &5:39:27.52 &35:41:30.2    &18.30 .08&18.35 .03&          &17.54 .02&         &17.23 .08&17.21 .02&           \\[-.1mm]
62    &      &5:39:27.53 &35:40:34.6    &         &         &          &         &         &17.34 .10&17.76 .06&17.41 .15  \\[-.1mm]
63    &      &5:39:27.58 &35:41:10.0    &         &         &          &         &         &         &18.76 .06&           \\[-.1mm]
64    &      &5:39:27.64 &35:40:06.3    &         &         &          &         &         &         &18.16 .07&           \\[-.1mm]
65    &      &5:39:27.68 &35:40:28.0    &         &         &          &         &         &         &18.01 .07&           \\[-.1mm]
66    &      &5:39:27.69 &35:40:09.1    &         &         &          &         &         &         &19.25 .12&           \\[-.1mm]
67    &10    &5:39:27.88 &35:40:42.3    &13.43 .03&13.42 .01&13.46  .03&13.06 .02&13.04 .03&12.91 .02&12.94 .01&12.86 .02  \\[-.1mm]
68    &      &5:39:27.89 &35:40:53.7    &         &         &          &         &         &16.97 .10&17.13 .04&16.74 .06  \\[-.1mm]
69    &21    &5:39:28.00 &35:40:51.9    &15.37 .01&15.37 .01&15.39  .04&14.75 .03&14.77 .04&14.46 .03&14.44 .02&14.43 .04  \\[-.1mm]
70    &45    &5:39:28.22 &35:42:00.1    &         &17.85 .08&          &16.08 .04&16.12 .06&         &15.17 .02&15.17 .06  \\[-.1mm]
71a   &      &5:39:28.30 &35:40:40.0    &16.93 .05&16.94 .03&16.97  .10&15.76 .03&15.80 .04&14.91 .05&14.77 .01&14.78 .04  \\[-.7mm]
71b   &      &5:39:28.29 &35:40:38.8    &         &19.62 .07&          &17.45 .02&17.41 .07&15.78 .05&15.77 .02&15.68 .05  \\[-.7mm]
{\bf 71}& &\multicolumn{2}{c}{\bf (a+b)}&16.80 .05&16.81 .04&16.88  .06&15.45 .05&15.45 .06&14.38 .05&14.31 .03&14.24 .04  \\[-.1mm]
72    &      &5:39:28.29 &35:40:44.9    &         &         &          &19.10 .32&         &14.70 .05&14.80 .01&14.56 .04  \\[-.1mm]
73    &      &5:39:28.31 &35:41:04.7    &         &         &          &         &         &17.56 .10&17.37 .02&           \\[-.1mm]
74    &      &5:39:28.35 &35:40:49.4    &         &         &          &         &         &         &18.06 .04&           \\[-.1mm]
75    &      &5:39:28.52 &35:41:06.8    &         &         &          &19.40 .07&         &17.45 .10&17.15 .02&           \\[-.1mm]
76a   &23    &5:39:28.50 &35:40:36.3    &         &         &          &18.45 .07&         &         &16.15 .08&           \\[-.7mm]
76b   &23    &5:39:28.54 &35:40:36.1    &         &         &          &18.57 .07&         &         &16.17 .07&           \\[-.7mm]
76c   &23    &5:39:28.57 &35:40:35.7    &         &         &          &18.72 .07&         &         &16.22 .07&           \\[-.7mm]
{\bf 76}&23 &\multicolumn{2}{c}{\bf (a+b+c)+nebula}&17.30 .10&18.65 .15&         &16.47 .15 &detected&13.59 .05&14.04 .10&14.71 .04  \\[-.1mm]
77    &      &5:39:28.61 &35:41:17.8    &detected &19.90 .15&          &17.51 .10&17.61 .08&15.72 .03&15.73 .02&15.73 .03  \\[-.1mm]
78a   &22    &5:39:28.61 &35:40:38.7    &         &detected         &          &17.75 .10&         &         &15.00 .03&  \\[-.7mm]
78b   &22    &5:39:28.58 &35:40:39.0    &         &18.41 .20&          &17.04 .10&         &         &15.55 .03&  \\[-.7mm]
78c   &22    &5:39:28.64 &35:40:39.2    &         &detected &          &17.72 .10&         &         &16.57 .08&  \\[-.7mm]
{\bf 78}&22&\multicolumn{2}{c}{\bf (a+b+c)+nebula}&17.50 .10&17.72  .15&detected &15.80 .05&15.90 .06&13.77 .04&14.06 .10&14.29 .10  \\[-.1mm]
79    &      &5:39:28.70 &35:41:47.3    &         &         &          &18.71 .06&         &17.89 .10&17.94 .06&           \\[-.1mm]
80    &      &5:39:28.87 &35:40:55.2    &         &         &          &19.58 .08&         &         &18.46 .06&           \\[-.1mm]
81    &      &5:39:28.91 &35:40:03.7    &         &         &          &19.69 .14&         &17.92 .10&18.05 .05&           \\[-.1mm]
82    &      &5:39:29.01 &35:41:19.6    &         &         &          &20.04 .15&        &         &18.70 .07&           \\[-.1mm]
83    &      &5:39:29.05 &35:40:05.2    &         &         &          &         &         &         &19.17 .17&           \\[-.1mm]
84    &      &5:39:29.13 &35:40:31.4    &         &         &          &         &         &17.74 .10&17.53 .05&           \\[-.1mm]
85    &11    &5:39:29.17 &35:41:39.9    &11.86 .02&11.94 .01&11.91  .02&11.71 .01&11.71 .02&11.61 .00&11.62 .01&11.61 .02  \\[-.1mm]
86    &      &5:39:29.20 &35:41:44.4    &         &18.53 .04&          &17.04 .04&17.02 .12&16.44 .08&16.36 .02&16.33 .08  \\[-.1mm]
87    &      &5:39:29.28 &35:40:46.8    &         &         &          &         &         &         &18.33 .05&           \\[-.1mm]
88    &      &5:39:29.30 &35:39:56.0    &19.29 .12&19.25 .11&          &17.90 .03&         &16.99 .10&17.06 .10&17.01 .10  \\[-.1mm]
89    &      &5:39:29.30 &35:40:02.7    &19.36 .12&19.31 .09&          &17.58 .04&17.72 .12&16.70 .04&16.75 .08&16.67 .08  \\[-.1mm]
90    &12    &5:39:29.33 &35:41:10.0    &detected &18.73 .04&          &15.16 .02&16.09 .03&14.01 .02&12.26 .01&13.23 .02  \\[-.1mm]
91    &      &5:39:29.52 &35:40:12.8    &         &         &          &18.25 .03&18.36 .12&16.34 .05&16.30 .01&16.41 .05  \\[-.1mm]
92    &      &5:39:29.52 &35:40:40.3    &         &         &          &         &         &         &19.01 .11&           \\[-.1mm]
93    &      &5:39:29.64 &35:41:52.4    &         &20.31 .20&          &18.74 .07&         &         &18.01 .06&           \\[-.1mm]
94    &25    &5:39:29.78 &35:41:31.1    &17.17 .01&17.20 .05&17.34  .12&16.39 .01&16.66 .05&16.10 .04&16.05 .05&16.11 .05  \\[-.1mm]
95    &35    &5:39:29.79 &35:39:24.2    &         &         &16.69  .20&         &16.09 .07&         &         &15.84 .06  \\[-.1mm]
96    &34    &5:39:29.89 &35:40:15.9    &17.35 .06&17.28 .05&17.52  .15&16.38 .03&16.58 .06&15.93 .03&15.90 .02&15.94 .06  \\[-.1mm]
97    &      &5:39:30.10 &35:41:52.7    &         &19.80 .14&          &18.33 .04&         &         &17.70 .04&           \\[-.1mm]
98    &33    &5:39:30.36 &35:40:15.1    &16.82 .05&16.78 .02&17.06  .15&15.81 .01&15.93 .02&15.27 .03&15.33 .02&15.28 .04  \\[-.1mm]
99    &      &5:39:30.43 &35:39:24.3    &         &         &          &         &17.46 .10&         &         &15.54 .10  \\[-.1mm]
100   &      &5:39:30.61 &35:41:51.4    &         &         &          &18.62 .13&         &         &18.22 .06&           \\[-.1mm]
101   &32    &5:39:30.87 &35:40:16.7    &15.82 .03&15.79 .01&15.94  .09&15.24 .02&15.40 .03&15.02 .02&15.05 .02&15.06 .05  \\[-.1mm]
102   &      &5:39:30.99 &35:41:32.0    &         &         &          &17.82 .03&         &16.01 .04&15.97 .02&16.09 .08  \\[-.1mm]
103   &      &5:39:31.04 &35:40:04.6    &         &19.71 .14&          &19.35 .10&         &         &18.81 .13&           \\[-.1mm]
104   &      &5:39:31.04 &35:41:37.3    &         &         &          &18.86 .06&         &         &17.81 .04&           \\[-.1mm]
105   &      &5:39:31.06 &35:40:40.1    &19.07 .10&19.08 .06&          &18.59 .05&         &18.19 .08&18.20 .05&           \\[-.1mm]
106   &      &5:39:31.09 &35:40:42.5    &         &19.81 .12&          &19.02 .08&         &         &18.06 .05&           \\[-.1mm]
107   &      &5:39:31.15 &35:41:29.8    &         &         &          &19.40 .12&         &         &18.71 .09&           \\[-.1mm]
108   &      &5:39:31.33 &35:41:12.6    &         &         &          &         &         &         &19.15 .12&           \\[-.1mm]
109   &      &5:39:31.38 &35:41:11.8    &         &         &          &         &         &18.04 .07&18.02 .05&           \\[-.1mm]
110   &30    &5:39:31.50 &35:40:54.9    &16.51 .04&16.51 .04&16.55  .10&15.85 .04&15.99 .05&15.63 .03&15.57 .01&15.60 .04  \\[-.1mm]
111   &      &5:39:31.60 &35:41:44.5    &         &         &          &         &         &         &18.12 .08&           \\[-.1mm]
112   &31    &5:39:31.63 &35:40:11.2    &17.46 .08&17.43 .04&17.52  .10&15.73 .03&15.86 .05&14.87 .02&14.84 .01&14.86 .02  \\[-.1mm]
113   &      &5:39:31.65 &35:40:11.8    &         &19.43 .10&          &18.79 .06&         &18.46 .11&18.60 .08&           \\[-.1mm]
\hline
\end{tabular}
\end{minipage}
}
\end{table*}
\begin{table*}
{\small
\begin{minipage}{500mm}
\contcaption{}
\begin{tabular}{@{}llllllllllll}
\hline
No:   &SW02  &RA &Dec &\multicolumn{3}{c}{---------------- $J$ Band ---------------- }&\multicolumn{2}{c}{-------- $H$ Band --------}&\multicolumn{3}{c}{---------------- $K$ Band ----------------} \\
      &ID    &          &               &20011226   &20021106   &20030319    &20021022 &20030319   &20011226 &20021022 &20030319       \\
\hline
114   &29    &5:39:31.80 &35:41:02.4    &17.91 .03&17.87 .02&          &16.22 .01&16.41 .05&15.40 .03&15.33 .01&15.52 .04  \\[-.1mm]
115   &      &5:39:31.83 &35:41:46.7    &         &         &          &19.14 .11&         &         &18.55 .12&           \\[-.1mm]
116   &      &5:39:31.89 &35:40:08.3    &         &19.19 .10&          &17.92 .04&         &17.05 .05&17.06 .03&17.46 .20 \\[-.1mm]
117   &      &5:39:31.89 &35:41:15.3    &         &19.18 .07&          &17.66 .03&17.87 .10&16.92 .03&16.79 .02&17.02 .10  \\[-.1mm]
118   &      &5:39:31.91 &35:40:30.3    &         &19.29 .09&          &17.89 .03&18.09 .10&17.05 .05&17.01 .03&17.13 .13  \\[-.1mm]
119   &27    &5:39:31.93 &35:41:37.3    &14.30 .00&14.30 .00&14.31  .03&13.01 .00&13.08 .02&12.49 .02&12.48 .01&12.49 .01  \\[-.1mm]
120   &28    &5:39:32.13 &35:41:19.3    &15.00 .03&15.02 .02&15.04  .04&14.01 .02&14.00 .03&13.39 .03&13.35 .02&13.35 .03  \\[-.1mm]
121   &      &5:39:32.27 &35:40:43.8    &19.18 .10&19.22 .08&          &18.11 .10&         &17.57 .06&17.53 .04&           \\[-.1mm]
122   &43    &5:39:32.31 &35:40:57.5    &17.85 .02&17.83 .03&          &16.88 .03&17.07 .06&16.31 .04&16.25 .02&16.57 .10  \\[-.1mm]
123   &      &5:39:32.45 &35:40:14.4    &         &19.05 .08&          &17.76 .08&         &17.08 .08&16.93 .03&17.17 .14  \\[-.1mm]
124   &      &5:39:32.58 &35:41:49.5    &         &19.13 .10&          &17.79 .08&         &         &         &17.02 .11  \\[-.1mm]
125   &37    &5:39:32.93 &35:40:01.7    &         &         &17.99  .17&         &16.25 .07&         &         &15.52 .05  \\[-.1mm]
126   &      &5:39:33.60 &35:41:15.3    &         &         &          &         &16.95 .12&         &         &16.59 .07  \\[-.1mm]
127   &38    &5:39:33.63 &35:39:25.0    &         &         &16.72  .08&         &15.33 .06&         &         &14.87 .05  \\[-.1mm]
128   &      &5:39:33.77 &35:40:55.8    &         &         &          &         &         &         &         &17.08 .10  \\[-.1mm]
129   &40    &5:39:34.19 &35:40:13.1    &         &         &16.56  .07&         &16.04 .03&         &         &15.82 .06  \\[-.1mm]
130   &39    &5:39:34.20 &35:39:30.3    &         &         &16.93  .07&         &16.09 .04&         &         &15.71 .06  \\[-.1mm]
131   &41    &5:39:34.23 &35:40:28.3    &         &         &17.35  .10&         &16.91 .04&         &         &16.72 .08  \\[-.1mm]
132   &42    &5:39:34.25 &35:40:40.0    &         &         &16.81  .07&         &16.31 .03&         &         &16.13 .07  \\[-.1mm]
133   &      &5:39:34.61 &35:41:10.2    &         &         &          &         &17.57 .11&         &         &detected   \\[-.1mm]
134   &      &5:39:35.00 &35:42:00.9    &         &         &16.16  .06&         &15.60 .06&         &         &15.46 .06  \\[-.1mm]
\hline
\end{tabular}
\end{minipage}
}
\end{table*}

\bsp

\label{lastpage}

\end{document}